\documentclass[12pt,a4paper]{article}
\usepackage{jheppub}  
\usepackage{amssymb} 
\usepackage{amsmath}
\usepackage{mathtools}
\usepackage{amsfonts}    
\usepackage{dsfont}
\usepackage{pdfpages}
\usepackage{verbatim}
\usepackage{tensor}

\newcommand{\ba}{\begin{align}}

\newcommand{\be}{\begin{equation}}
\newcommand{\ee}{\end{equation}}
\def\bd{\begin{tikzpicture}}
\def\ed{\end{tikzpicture}}
\newcommand{\abs}[1]{\left| #1 \right|}
\newcommand{\ket}[1]{| #1 \rangle}

\allowdisplaybreaks[1]

\title{The plane-wave spectrum from the worldsheet}

\author{Lorenz Eberhardt, Kevin Ferreira} 
\affiliation{Institut f\"ur Theoretische Physik, ETH Zurich, \\
\hspace*{0.3cm}CH-8093 Z\"urich, Switzerland}
\emailAdd{eberhardtl@itp.phys.ethz.ch, kferreira@itp.phys.ethz.ch}

\abstract{We study string theory on $\mathrm{AdS}_3$ backgrounds with mixed flux using the hybrid formalism of Berkovits, Vafa and Witten. We solve the worldsheet description of the theory completely in the plane-wave limit. This constitutes a direct derivation of the plane-wave spectrum from the worldsheet with mixed flux.}

\begin{document}

\maketitle

%make math in all titles bold
\makeatletter
\g@addto@macro\bfseries{\boldmath}
\makeatother
%end code

%%%%%%%%%%%%%%%%%%%%%%%%%%%%%%%%%%%%%%%%%%%%%%%%%%%%%%%%%%%%%%%
\section{Introduction} \label{sec:intro}
%%%%%%%%%%%%%%%%%%%%%%%%%%%%%%%%%%%%%%%%%%%%%%%%%%%%%%%%%%%%%%%

String theory on the backgrounds $\mathrm{AdS}_3 \times \mathrm{S}^3 \times \mathbb{T}^4$ and $\mathrm{AdS}_3 \times \mathrm{S}^3 \times \mathrm{K3}$ plays an important r\^ole in the $\mathrm{AdS}_3/\mathrm{CFT}_2$-correspondence. 
The two backgrounds arise as a near-horizon limit of the D1-D5 system on $\mathbb{T}^4$ and $\mathrm{K3}$, and are conjectured to be dual to the symmetric orbifold CFT of $\mathbb{T}^4$ and $\mathrm{K3}$, respectively \cite{Maldacena:1997re, Aharony:1999ti}. 
The $\mathrm{AdS}_3/\mathrm{CFT}_2$-correspondence is particularly rich, since $\mathrm{AdS}_3$ can be supported by a mixture of NS-NS and R-R flux.
Combined with its accessibility, it constitutes an attractive setting to explore the fundamental aspects of holography and string theory.

While in the full string theory the fluxes are quantised, this is not so in a perturbative worldsheet description. 
Indeed, if $g$ denotes the string coupling constant, the worldsheet description holds in the limit $g \to 0$.
The number of R-R quanta must be of order $g^{-1}$ to have a visible effect on the background geometry.
Thus the R-R flux is essentially continuous and appears as a modulus in the worldsheet theory, as well as in the dual CFT. 

For pure NS-NS flux, the worldsheet theory admits a description in terms of an $\mathcal{N}=1$ supersymmetric WZW-model \cite{Henningson:1991jc, Maldacena:2000hw, Maldacena:2000kv, Maldacena:2001km, Giveon:1998ns, Kutasov:1999xu, deBoer:1998gyt} in the RNS-formalism, which is completely solvable. 
In particular, the string spectrum on $\mathrm{AdS}_3 \times \mathrm{S}^3 \times \mathcal{M}_4$, where $\mathcal{M}_4=\mathbb{T}^4$ or $\mathrm{K3}$, is known exactly. 
This is in sharp contrast with the situation with mixed flux. 
The worldsheet theory with mixed flux was worked out in \cite{Berkovits:1999im, Dolan:1999dc} in a Green-Schwarz-like formalism. 
This formulation is also called hybrid formalism and is reviewed in the beginning of Section~\ref{sec:applications}.

For the case with mixed flux, an integrability description of the theory exists \cite{Babichenko:2009dk, Borsato:2012ud, Hoare:2013lja, Lloyd:2014bsa, Sfondrini:2014via, Sundin:2014ema, OhlssonSax:2018hgc} in the decompactification limit \cite{Arutyunov:2009ga}, in which the worldsheet becomes a plane. 
This paper establishes a direct link between integrability computations and worldsheet computations.

The hybrid formalism features a principal chiral model on a supergroup with a WZW-term.
In the case of $\mathrm{AdS}_3 \times \mathrm{S}^3 \times \mathcal{M}_4$, this is a sigma-model on $\mathrm{PSU}(1,1|2)$. 
This principal chiral model admits two parameters: the normalisation of the kinetic term, and the coefficient of the WZW-term. 
In string theory, the normalisation of the kinetic term is related to the total flux, whereas the WZW-term is related to the NS-NS flux and its coefficient is therefore quantised. 
For the precise relation, see \eqref{eq:relation f and Qs}. 
In \cite{Gaberdiel:2011vf,Gerigk:2012cq}, the pure NS-NS flux case was treated from a supergroup perspective.
On a general group, only the theory with equal normalisations of the kinetic and the WZW-term is conformal. 
On supergroups with vanishing dual Coxeter number the theory is conformal for all values of the parameters. 
However, the chiral algebra of the theory becomes dramatically smaller away from the WZW-point. In particular, the theory is not rational and representation theory of the chiral algebra does not impose strong enough constraints to determine the spectrum of the theory.

To make progress in the computation of the spectrum of the CFT on such supergroups, we follow a different route initiated in \cite{Ashok:2009xx, Benichou:2010rk}. 
The model still has a global $\mathrm{PSU}(1,1|2) \times \mathrm{PSU}(1,1|2)$-symmetry, and one can work out the OPE's of the conserved currents exactly, thanks to a non-renormalisation theorem \cite{Bershadsky:1999hk,Quella:2007sg}. 
The currents are non-holomorphic, which complicates certain arguments. 
One finds that the OPE's of the currents close up to an additional field, which will be described in detail below. 
This algebra generates the spectrum of the CFT, in the same way as the affine Ka\v c-Moody algebra does for a WZW-model. 
However, several states of the spectrum mix under the action of $L_0$, and therefore $L_0$ needs to be diagonalised in order to obtain the conformal weights.

In this paper we solve the theory completely in a BMN-like limit, which fully reproduces the plane-wave spectrum of string theory on $\mathrm{AdS}_3 \times \mathrm{S}^3 \times \mathcal{M}_4$. 
This constitutes a direct contact between the worldsheet theory and the string spectrum on a background with R-R flux. 
The corresponding analysis in the pure NS-NS background was first carried out in \cite{Son:2003zv}. 
We repeat the analysis also for the background $\mathrm{AdS}_3 \times \mathrm{S}^3 \times \mathrm{S}^3 \times \mathrm{S}^1$, which supports the large ${\cal N}=(4,4)$ algebra and has received some interest recently. 
The relevant supergroup is in this case $\mathrm{D}(2,1;\alpha)$. 
We derive the full plane-wave spectrum on this background and, in particular, the masses of the fields in the GS-formalism agree with \cite{Rughoonauth:2012qd, Abbott:2012dd,Dei:2018yth}.

\medskip

This paper is organized as follows. In Section~\ref{sec:classical}, we present classical solutions of the sigma model on supergroups, and compute their energy. 
We subsequently start in Section~\ref{sec:current} with a quantum treatment of the model by employing the existence of a non-holomorphic current algebra. 
We define representations of this current algebra in Section~\ref{sec:representations}, and discuss its spectral flow symmetry. 
The algebra considerably simplifies in the limit of large charges, which is discussed in Section~\ref{sec:large charge}. 
This will confirm and generalise the result obtained via classical solutions in Section~\ref{sec:current}, by including the relevant quantum corrections. 
We apply these results to string theory in Section~\ref{sec:applications} and derive the BMN-formula. We discuss our findings in Section~\ref{sec:discussion}.

%%%%%%%%%%%%%%%%%%%%%%%%%%%%%%%%%%%%%%%%%%%%%%%%%%%%%%%%%%
\section{Semiclassical analysis}\label{sec:classical}
%%%%%%%%%%%%%%%%%%%%%%%%%%%%%%%%%%%%%%%%%%%%%%%%%%%%%%%%%%

In this section we perform a semiclassical analysis of the worldsheet sigma-model that will be our focus throughout the paper.
We will do this by finding the worldsheet conformal weight of some classical solutions, and interpreting them semiclassically.

%%%%%%%%%%%%%%%%%%%%%%%%%%%%%%%%%%%%%%%%%%%%%%%%%%%%%%%%
\subsection{Classical action and conserved currents}
%%%%%%%%%%%%%%%%%%%%%%%%%%%%%%%%%%%%%%%%%%%%%%%%%%%%%%%%
We consider the two-parameter sigma model on a (super)group G
\begin{equation}\label{eq:action}
\mathcal{S}[g] = -\frac{1}{4\pi f^2} \int d^2z \, \text{Tr}\left( \partial g g^{-1}\, \bar{\partial} g g^{-1}\right) + k\, \mathcal{S}_{\text{WZ}}[g] \, ,
\end{equation}
with $g\in{\rm G}$, and where $\mathcal{S}_{\text{WZ}}[g]$ denotes the Wess-Zumino term.
The points $kf^2=\pm 1$ on parameter space correspond to the usual WZW-model.
At these points \eqref{eq:action} possesses a local ${\rm G} \times {\rm G}$ symmetry, while at $k=0$ we recover the principal chiral model \cite{Bershadsky:1999hk}.

Away from the WZW-point, the model still has a global ${\rm G} \times {\rm G}$ symmetry, which gives rise to two local conserved currents.\footnote{
There is another deformation of the WZW-model which preserves conformal symmetry and gives rise to a current algebra, see \cite{Konechny:2010nq}. However, it only preserves the diagonal global ${\rm G}$ symmetry.} 
Let us focus on the symmetry by left-multiplication, with associated current $j(z,\bar{z})$.\footnote{ 
The right-multiplication can be treated similarly, and in fact the components of its associated current $\bar{j}(z,\bar{z})$ are conjugate to those of $j(z,\bar{z})$. 
In particular, they will give rise to the same energy-momentum tensor.
}
This current has the following components in complex coordinates:
\be \label{eq:currrents def}
j_z=-\frac{1+kf^2}{2f^2} \partial gg^{-1}\ , \quad j_{\overline{z}}=-\frac{1-kf^2}{2f^2} \overline{\partial}gg^{-1}\, ,
\ee
and the equations of motion of \eqref{eq:action} are equivalent to the conservation law
\begin{equation} \label{eq:eom}
\bar{\partial}j_z + \partial j_{\bar{z}} = 0 \, .
\end{equation}
At the WZW-point $kf^2=1$, $j_{\bar{z}}\equiv 0$, and conservation implies the holomorphicity of the $j_z$ component. However, we stress that in general $j_z$ is neither holomorphic nor anti-holomorphic.
We will henceforth write $j_z(z)$ but it is understood that no quantity is assumed to be purely holomorphic or anti-holomorphic.
The associated Noether charges are given by the integral of the time component of the currents over a constant time slice
\be \label{eq:charges}
Q \equiv \oint\limits_{\abs{z}=R} \frac{\text{d}z}{z}j^{t}(z)=\oint\limits_{\abs{z}=R} \text{d}z \bigg(j_z(z)+\frac{\bar{z}}{z}j_{\bar{z}}(z)\bigg)\, .
\ee
Note that, in contrast with the usual techniques in CFT, the integration contour cannot be deformed since the currents are not holomorphic.
The conservation of the current ensures the independence of the charge on the radius $R$, so we will fix $R\equiv 1$ from here on.
The action \eqref{eq:action} has conformal symmetry for any values of $k$ and $f^2$ since the energy-momentum tensor is holomorphic:
\be  \label{eq:energy_mom}
T(z)=\frac{2f^2}{(1+kf^2)^2} \, \text{Tr}\left( j_z(z) j_z(z) \right) = \frac{2f^2}{(1-kf^2)^2} \, \text{Tr}\left( \bar{j}_z(z) \bar{j}_z(z)\right)\, ,
\ee
where the second equality follows from the existing conjugacy relation between the components of the left and right currents.
Throughout this paper $L_n$ will denote the modes of the expansion of the energy-momentum tensor, as usual.
If the dual Coxeter number of G vanishes (as for ${\rm PSU}(1,1\vert 2)$), several non-renormalisation theorems on two- and three-point functions ensure that this symmetry is preserved at the quantum level \cite{Berkovits:1999im, Bershadsky:1999hk, Quella:2007sg, Benichou:2010rk}.

%%%%%%%%%%%%%%%%%%%%%%%%%%%%%%%%%%%%%%%%%%%%%%%
\subsection{Ground state solutions}
%%%%%%%%%%%%%%%%%%%%%%%%%%%%%%%%%%%%%%%%%%%%%%%

The classical ground state solution is given by
\begin{equation}\label{eq:classical ground state}
g(z,\bar{z})=\exp\left( - k f^2 \, \log (z\bar{z}) \,  (\nu \cdot t)\right) \, ,
\end{equation}
where $t^i$ is an element of the Cartan subalgebra of $\mathfrak{g}$, the Lie algebra of G, and $i=1,\ldots,{\rm rank}(\mathfrak{g})$.
Furthermore $\exp$ denotes the Lie algebra exponential. We use here the inner product $\nu \cdot t\equiv \nu^i \kappa_{ij} t^j$, where $\kappa_{ij}$ is an invariant form on the Lie algebra.
It can easily be shown that this solution indeed satisfies the equations of motion.
The conserved charges are given by $Q=\bar{Q}= k\nu^i \kappa_{ij} t^j$, so that $\nu^i$ is interpreted semiclassically as $\nu^i = \ell^i_0/k$, where $Q^i=\bar{Q}^i=\ell^i_0$ are the charges of the ground state solution.
Finally, the energy-momentum tensor is
\begin{equation}\label{eq:ground_state_L0}
T(z) = \frac{f^2 (\ell_0\cdot \ell_0)}{2z^2} \, ,
\end{equation}
and similarly for $\bar{T}(\bar{z})$.
In Section \ref{sec:applications} we will find the quantum correction to this result.

%%%%%%%%%%%%%%%%%%%%%%%%%%%%%%%%%%%%%%%%%
\subsection{Excited solutions}
%%%%%%%%%%%%%%%%%%%%%%%%%%%%%%%%%%%%%%%%%

Consider now the following excited solution:%\footnote{The general form of the excitation is fixed by a reality condition on $g(z,\bar{z})$.}
\begin{equation}\label{eq:classical excitation}
g(z,\bar{z})=\exp\left(
   \frac{1}{\sqrt{k}} \left( \mu \, z^{\alpha} \bar{z}^{\beta} t^a - 
      \mu^* \,  z^{-\alpha}\bar{z}^{-\beta} t^{-a}\right)\right)\, \exp\left(-kf^2 \,  \log(z \bar{z})\,(\nu\cdot t) \right) \, ,
\end{equation}
where $t^a$ is a step operator or a Cartan-element of $\mathfrak{g}$, and $\mu$ is a coefficient to be fixed.
Furthermore, $t^{-a}$ denotes the step operator associated with the opposite root. 
This has to be included to ensure the reality of the solution.
Single-valuedness requires $\beta-\alpha = n$, with $n$ an integer.
Finally, the equations of motion \eqref{eq:eom} are obeyed provided that\footnote{For this solution we have chosen a specific branch of the square-root, by assuming that $n\geq 2\nu$.
The other branch can be obtained from the first by considering $n\leq 2\nu$.
}
\begin{equation}
\alpha=\frac{1}{2} \left(-n - (a\cdot \nu)\,  kf^2 + \sqrt{n^2-2 (a\cdot\nu)\,  n k^2f^4 +  (a\cdot\nu)^2 k^2f^4}\right)  \, ,
\end{equation}
where $a^i$ denotes $\alpha^i$ if $a$ is a root, and $0$ if $a$ is a Cartan-index. 
Plugging \eqref{eq:classical excitation} into \eqref{eq:charges} and using \eqref{eq:currrents def}, the charges associated with these excited solutions can be explicitly computed.
The expressions for these charges in terms of $\mu$, $\nu$, $n$ are quite involved, so we will not reproduce them here.
Instead, we will use the parametrisation

\begin{equation}\label{eq:charge parametrisation}
Q = \ell_0 \cdot t+N_n (a \cdot t)\, , \qquad \bar{Q} = \ell_0 \cdot t \, ,
\end{equation}
which allows us to trade $\nu^i$ for $\ell_0^i$ and $\mu$ for $N_n$.
The plane-wave limit may now be obtained by considering $n, N_n\ll k,~\ell_0^i,~f^{-2}$.
In this limit the zero-mode of the energy-momentum tensor \eqref{eq:energy_mom} becomes
\begin{equation}
L_0 = \frac{f^2(\ell_0 \cdot \ell_0)}{2} + \frac{N_n}{2}\left(n + (a\cdot \ell_0) f^2+\sqrt{n^2 - 2 (a\cdot\ell_0)  n kf^4 + (a\cdot\ell_0)^2f^4}\right) + \mathcal{O}\big(k^{-1}\big)\, .
\end{equation}
This expression is to be compared with \eqref{eq:BMN conformal weight}, which will be obtained from the full quantum treatment developed in the following sections.
In Section \ref{sec:applications} these results will be applied to string theory, and the full BMN formula of \cite{Berenstein:2002jq} will be obtained.
This preliminary result shows that worldsheet methods based on \eqref{eq:action} may give us access to the plane-wave spectrum with mixed flux. %while enhancing the classical character of the plane-wave limit.

Semiclassically, the parametrisation \eqref{eq:charge parametrisation} suggests that this state is obtained from the groundstate of charge $\ell_0$ by the application of $N_n$ generators of the left symmetry, with mode number $n$.
This interpretation is reinforced by the observation that, to all orders in $k^{-1}$,
\begin{equation}
L_0-\bar{L}_0 = n N_n \, ,
\end{equation}
which is indeed quantised in the quantum theory.

\medskip

Finally, we stress that this is an exact classical solution and is hence expected to yield the correct conformal weight in the classical limit. The classical limit is given by $k,~\ell_0^i,~f^{-2}\rightarrow\infty$, with all their ratios fixed, and for any $n$ and $N_n$. In particular, this is a much more powerful limit than the plane-wave limit, and even the decompactification limit \cite{Arutyunov:2009ga}. However, we are so far limited in our computations, in that we have only managed to find a single-excitation solution.\footnote{In particular our solution will not be level-matched in string theory.}

%%%%%%%%%%%%%%%%%%%%%%%%%%%%%%%%%%%%%%%%%%%%%%%%%%%%%%%%%%%%%%%
\section{Review of the current algebra} \label{sec:current}
%%%%%%%%%%%%%%%%%%%%%%%%%%%%%%%%%%%%%%%%%%%%%%%%%%%%%%%%%%%%%%%

In this section we review the current algebra introduced in \cite{Ashok:2009xx} and further analysed in \cite{Benichou:2010rk, Ashok:2009jw}, which will be the main tool of this work. 
In particular, its applications to string theory via the hybrid formalism \cite{Berkovits:1999im} will be described in Section \ref{sec:applications}.

%%%%%%%%%%%%%%%%%%%%%%%%%%%%%%%%%%%%%%%%%
\subsection{Conformal current algebra}
%%%%%%%%%%%%%%%%%%%%%%%%%%%%%%%%%%%%%%%%%

A non-chiral current algebra in two-dimensions compatible with conformal symmetry was first formulated in \cite{Ashok:2009xx} in all generality.
This algebra was constructed at the level of OPE's by requiring their consistency with locality, and Lorentz and parity-time reversal symmetries.
The non-linear sigma models of the kind \eqref{eq:action} were then seen to consistently realise the constructed general OPE structure, by computing the current-current correlators and the OPE's of the models in conformal perturbation theory. 
This result holds for sigma models based on Lie supergroups whose superalgebra has vanishing Killing form, such as PSU($1,1\vert 2)$ (see Appendix \ref{app:properties} for a review of the relevant properties of Lie superalgebras). 
For those, a non-renormalization theorem \cite{Bershadsky:1999hk} allows one to do the computation to all orders in perturbation theory.

\medskip

The OPE's between the components of the currents were found to be as follows:
\begin{align}
j^{a}_z(z)j^{b}_z(w)&\sim \frac{(1+kf^2)^2\kappa^{ab}}{4f^2(z-w)^2}+\frac{\mathrm{i}}{4}\tensor{f}{^{ab}_c}\bigg(\frac{(3-kf^2)(1+kf^2)}{z-w}\, j^c_z(w)\nonumber\\
&\qquad\qquad\qquad+\frac{(1-kf^2)^2(\bar{z}-\bar{w})}{(z-w)^2} \, j^c_{\bar{z}}(w)\bigg)\, , \label{eq:current_algebra1}\\
j^{a}_{\bar{z}}(z)j^{b}_{\bar{z}}(w)&\sim \frac{(1-kf^2)^2\kappa^{ab}}{4f^2(\bar{z}-\bar{w})^2}+\frac{\mathrm{i}}{4}\tensor{f}{^{ab}_c}\bigg(\frac{(3+kf^2)(1-kf^2)}{\bar{z}-\bar{w}}\, j^{c}_{\bar{z}}(w)\nonumber\\
&\qquad\qquad\qquad+\frac{(1-kf^2)^2(z-w)}{(\bar{z}-\bar{w})^2}\, j^{c}_z(w)\bigg) \, , \label{eq:current_algebra2}\\
j^a_z(z)j^b_{\bar{z}}(w) &\sim % -\frac{2\pi(kf^2+1)(kf^2-1)\kappa^{ab}}{4f^2} \delta^{(2)}(z-w)
(1-kf^2)^2\tensor{f}{^{ab}_c}\bigg(\frac{j^{c}_z(w)}{(\bar{z}-\bar{w})} +\frac{j^c_{\bar{z}}(w)}{(z-w)} \bigg)\, ,\label{eq:current_algebra3}
\end{align}
where $\sim$ denotes equality up to regular and contact terms. A regular term is by definition less divergent than a pole, in particular there are logarithmic corrections to these OPE's. Their explicit form can be found in \cite{Benichou:2010rk}.
Here $\kappa^{ab}$ and $\tensor{f}{^{ab}_c}$ are the components of the invariant tensor and the structure constants of $\mathfrak{g}$, respectively (see Appendix \ref{app:properties}).
Notice that at the WZW-point this current algebra reduces to a Ka\v c-Moody algebra.% (in particular, all logarithmic corrections vanish).

%%%%%%%%%%%%%%%%%%%%%%%%%%%%%%%%%%%%%%%%
\subsection{Energy-momentum tensor}
%%%%%%%%%%%%%%%%%%%%%%%%%%%%%%%%%%%%%%%%

The holomorphic energy-momentum tensor is as usual the regularisation of its classical counterpart:
\be \label{eq:energy_mom_quantum}
T(z)=\frac{2f^2}{(1+kf^2)^2} \, \kappa_{ab} (j^a_z j^b_z)(z) = \frac{2f^2}{(1-kf^2)^2} \, \kappa_{\bar{a}\bar{b}} (\bar{j}^{\bar{a}}_z \bar{j}^{\bar{b}}_z)(z)\, ,
\ee
It was shown in \cite{Benichou:2010rk} that this energy-momentum tensor is indeed holomorphic. In fact,
\be 
W^{(s)}(z)=d_{a_1 \cdots a_s} (j^{a_1}_z \cdots j^{a_s}_z)(z)
\ee
is holomorphic for every Casimir $d_{a_1 \cdots a_s}t^{a_1} \cdots t^{a_s}$ of $\mathfrak{g}$. These fields generate the full chiral algebra of the CFT.\footnote{The algebra $\mathfrak{psu}(1,1\vert 2)$ possesses one further Casimir of order 6 for which the result applies, so the chiral algebra of this theory is a $\mathcal{W}(2,6)$-algebra. For an explicit construction of this algebra, see \cite{Blumenhagen:1990jv}.} This chiral algebra is much too small to constrain the spectrum of the CFT and is hence not very useful for our purpose. In particular, the CFT is not rational.

Despite the fact that $j_z(z)$ and $j_{\bar{z}}(z)$ are not holomorphic nor anti-holomorphic, their OPE's with $T(z)$ are those of primary fields of dimension one and zero, respectively:
\begin{align}
\begin{aligned}
T(z)j_z^a(w) &\sim \frac{j_z^a(w)}{(z-w)^2}+ \frac{\partial j_z^a(w)}{z-w}\, , \\
T(z)j_{\bar{z}}^a(w) &\sim \frac{\partial j_{\bar{z}}^a(w)}{z-w}\, ,
\end{aligned}
\end{align} 
possibly with logarithmic corrections.
We take this as an indication that it is useful to think of the currents and their OPE's as the spectrum-generating algebra, even away from the WZW-point. We will see in Section~\ref{sec:large charge} that this is true in a BMN-like limit.

%%%%%%%%%%%%%%%%%%%%%%%%%%%%%%%%%%%%%%%%%%%%%%%%%%%%%%%%%%%%%%%%%%%%%%
\subsection{Conserved charges and the mode algebra}\label{sec:modes}
%%%%%%%%%%%%%%%%%%%%%%%%%%%%%%%%%%%%%%%%%%%%%%%%%%%%%%%%%%%%%%%%%%%%%%

As usual in quantum field theory, the symmetry algebra must be realised on the Hilbert space of the theory through a set of conserved charges.\footnote{The symmetry could also be anomalous, but we will see shortly that this is not the case.} 
These charges were introduced in \eqref{eq:charges}, and their bracket $[Q^a,Q^b]$ may now be computed. 
Here and in the following it is implicit that if both $a$ and $b$ are fermionic indices the bracket $[Q^a,Q^b]$ is to be understood as an anti-commutator.
Moreover, for the sake of simplicity we suppress possible signs arising from the fermionic nature of the supercurrents.
Nevertheless, our final results hold for bosonic as well as for fermionic currents.
The computation is subtle since we cannot rely on usual CFT techniques like contour deformation. 
However the commutator can be written as
\begin{multline}
[Q^a,Q^b]=\lim_{\epsilon \downarrow 0} \Bigg(\ \oint\limits_{\abs{z}=R+\epsilon} \!\!\!\mathrm{d}z \ \oint\limits_{\abs{z}=R}\!\mathrm{d}w - \oint\limits_{\abs{z}=R-\epsilon}\!\!\! \mathrm{d}z \ \oint\limits_{\abs{z}=R} \!\mathrm{d}w \Bigg) \\
\times\left(j_z^a(z)+\frac{\bar{z}}{z}j_{\bar{z}}^a(z)\right)\left(j_z^a(w)+\frac{\bar{w}}{w}j_{\bar{z}}^a(w)\right)\, .
\end{multline}
Inserting the OPE's \eqref{eq:current_algebra1}--\eqref{eq:current_algebra3} and performing the integrals we indeed obtain
\be 
[Q^a,Q^b]=\mathrm{i}\tensor{f}{^{ab}_c} Q^c\, .
\ee
This is a very good consistency check on the construction.
%Note that in this computation the logarithmic contributions and regular contributions of the OPE's were not considered, since the currents are not holomorphic they can in principle also give a contribution. However, we expect this contribution to be absent.
Similarly, one can compute the commutators of $Q^a$ with the modes of the energy-momentum tensor $L_n$ and $\bar{L}_n$. 
Holomorphicity of $T(z)$ simplifies the computation considerably, and yields the expected result
\be 
[L_n,Q^a]=[\bar{L}_n,Q^a]=0\, ,
\ee
i.e.~the internal and conformal symmetries commute. 
In particular, this shows that the charge is indeed conserved, since it commutes with the Hamiltonian $L_0+\bar{L}_0$.

\medskip

Motivated by this construction, we now define a convenient set of operators (of which the conserved charges above form a subset) which allows us to build the spectrum of our model.
In analogy with the usual chiral currents in CFT we define\footnote{In contrast with the usual conventions in CFT, all operators are defined via contour $z$-integrals.
The contour relation $z=R^2 \bar{z}^{-1}$ will lead to some unusual signs in our modes.
On the other hand, in line with usual QFT results, no physical implication stems from the actual value of $R$.}
\begin{align}\label{eq:X_Y}
X^a_n \equiv \oint\limits_{\abs{z}=R} \frac{\mathrm{d}z}{R} \, z^n j_z^a(z) \, ,\qquad 
Y^a_n \equiv \oint\limits_{\abs{z}=R}  \frac{\mathrm{d}z}{R} \, z^{n-1}\bar{z}\, j_{\bar{z}}^a(z)\,  ,
\end{align}
and analogously for the right-current $\bar{j}(z)$, which give rise to operators $\bar{X}_n^a$, $\bar{Y}_n^a$.
As before, the commutation relations of these quantities can be worked out. 
It is quite convenient to define the combinations
\begin{align}\label{eq: Q and P defs}
\begin{aligned}
Q_n^a & = X_n^a+Y_n^a
\, , & P_n^a&= 2kf^2 \left( \frac{X_n^a}{1+kf^2} - \frac{Y_n^a}{1-kf^2}\right)\, , \\
\bar{Q}_n^{\bar{a}}&=\bar{X}_n^{\bar{a}}+\bar{Y}_n^{\bar{a}}
\, ,& \bar{P}_n^{\bar{a}}&=-2kf^2 \left( \frac{\bar{X}_n^{\bar{a}}}{1-kf^2} - \frac{\bar{Y}_n^{\bar{a}}}{1+kf^2}\right)\, ,
\end{aligned}
\end{align}
for which we find the commutation relations
\begin{align}\label{eq:mode_algebra}
\begin{aligned}
\,[Q_m^a,Q_n^b] & = k m \kappa^{ab} \delta_{m+n,0}+\mathrm{i} \tensor{f}{^{ab}_c} Q^c_{m+n}
\ ,& [Q_m^a,\bar{P}_n^{\bar{b}}]&=k m A^{a\bar{b}}_{m+n}\, , \\
[Q_m^a,P_n^b] & = k m\kappa^{ab}\delta_{m+n,0}+\mathrm{i} \tensor{f}{^{ab}_c} P_{m+n}^c 
\, ,& [\bar{Q}_m^{\bar{a}},A^{b\bar{b}}_n]&=\mathrm{i}\tensor{f}{^{\bar{a}\bar{b}}_{\bar{c}}} A^{b\bar{c}}_{m+n}\, , \\
[\bar{Q}_m^{\bar{a}},\bar{Q}_n^{\bar{b}}]&=-k m\kappa^{ab} \delta_{m+n,0}+\mathrm{i} \tensor{f}{^{\bar{a}\bar{b}}_{\bar{c}}} Q^{\bar{c}}_{m+n}
\, ,&
[Q_m^a,A^{b\bar{b}}_n]&=\mathrm{i}\tensor{f}{^{ab}_c} A^{c\bar{b}}_{m+n}\, , \\
[\bar{Q}_m^{\bar{a}},\bar{P}_n^{\bar{b}}]&= -k m\kappa^{\bar{a}\bar{b}}\delta_{m+n,0}+\mathrm{i} \tensor{f}{^{\bar{a}\bar{b}}_{\bar{c}}} \bar{P}_{m+n}^{\bar{c}}
\, ,&
[\bar{Q}_m^{\bar{b}},P_n^a]&=-km A^{a\bar{b}}_{m+n}\, ,
\end{aligned}
\end{align}
with all other commutators vanishing, and in particular $[P_m^a,P_n^b]=0$. 
The barred and unbarred modes constitute two non-commuting non-semisimple super-Ka\v c-Moody algebras at level $k$ and $-k$.\footnote{The negative sign of one of the levels is immaterial: it is simply a consequence of our unusual conventions for barred modes.}
A rescaled version of this algebra appears already in \cite{Ashok:2009xx}.
The non-commutativity of these algebras is of course related to the non-holomorphicity of the currents, and it is encoded in the bi-adjoint field
\begin{equation}\label{eq: A definition}
A^{a\bar{a}} = \text{STr}\left( g^{-1} t^a g \, t^{\bar{a}}\right) \, ,
\end{equation}
where $t^a$, $t^{\bar{a}}$ are the generators of each of the two copies of $\mathfrak{g}$ in the adjoint representation. 
It has conformal weight zero, since the Casimir of the adjoint representation of $\mathfrak{g}$ vanishes.

Since $P_n^a$ commutes with itself, its scaling is arbitrary.\footnote{Note that $P=-k\text{Tr}\left(g^{-1}\partial_\phi g\right)$, where $\phi$ is the compact direction on the worldsheet.
It is then natural that $P$ commutes with itself, since it contains no time derivatives.} 
Therefore the only meaningful parameter which appears and which is subject to possible unitarity restrictions is $k$. 
%It thus looks like the Fock space spanned by these modes is unitary (after dividing out the null space). 
At the WZW-point, the $Y_n^a$ become null fields and $Q_n^a$ reduce to the modes of the chiral currents of the WZW model.\footnote{Analogously, at the WZW-point $\bar{X}_n^a$ become null and $\bar{Q}_n^a$ reduce to modes of the anti-chiral currents.}
Finally, note that the conserved charges constructed above are simply the zero-modes $Q^a_0$.

\medskip

It is important to notice that not all the modes of the mode algebra are independent, i.e.~this mode algebra does not act faithfully on the Hilbert space.
The relations between the different modes can be found in \cite{Benichou:2010rk}, and their precise form is mostly irrelevant for our results.

%%%%%%%%%%%%%%%%%%%%%%%%%%%%%%%%%%%%%
\subsection{The Virasoro modes}
%%%%%%%%%%%%%%%%%%%%%%%%%%%%%%%%%%%%%

Since $T(z)$ is holomorphic, the computation of the commutation relations of the Virasoro modes with the current modes can be simplified by contour-deformation techniques, and by ignoring non-singular terms in the OPE's.
%Nevertheless, the computation remains quite lengthy and cumbersome.
Alternatively, we can use \eqref{eq:energy_mom} and \eqref{eq:X_Y} to first write
\begin{equation}\label{eq:Virasoro_modes}
L_n = \frac{2f^2}{(1+kf^2)^2}\kappa_{ab}\big( X^aX^b\big)_n = \frac{2f^2}{(1-kf^2)^2}\kappa_{\bar{a}\bar{b}}\big( \bar{X}^{\bar{a}}\bar{X}^{\bar{b}}\big)_n \, .
\end{equation}
and then take commutators of normal-ordered products as usual. 
The two methods yield the same result, namely the following commutation relations:
\begin{align}\label{eq:Virasoro_commutations}
\begin{aligned}
\,[L_m,Q_n^a] & = -  \frac{1+kf^2}{2}n Q_{n+m}^a - \frac{1-k^2f^4}{4kf^2}nP_{m+n}^a  \, , \\
[L_m,P_n^a] & = -  kf^2nQ^a_{n+m} - \frac{1-kf^2}{2}nP_{n+m}^a - \text{i}f^2\tensor{f}{^a_{bc}}\big( Q^bP^c\big)_{n+m} \, ,\\
[L_m,\bar{Q}_n^{\bar{a}}] & = -  \frac{1-kf^2}{2} n\bar{Q}_{n+m}^{\bar{a}} + \frac{1-k^2f^4}{4kf^2} n\bar{P}_{m+n}^{\bar{a}} \, , \\
[L_m,\bar{P}_n^{\bar{a}}] & =  kf^2 n\bar{Q}^{\bar{a}}_{n+m} -  \frac{1+kf^2}{2}n\bar{P}_{n+m}^{\bar{a}} - \text{i}f^2\tensor{f}{^{\bar{a}}_{\bar{b}\bar{c}}}\big( \bar{Q}^{\bar{b}}\bar{P}^{\bar{c}}\big)_{n+m} \, .
\end{aligned}
\end{align}
These results can be derived from both expressions for the Virasoro modes in \eqref{eq:Virasoro_modes}. 
Note that the result above is independent of the normal-ordering scheme we use, since $\tensor{f}{^a_{bc}}Q^b_m P^c_n=\tensor{f}{^a_{bc}} P^c_m Q^b_m$ because of the vanishing of the dual Coxeter number.
It is important to notice that, due to the appearance of normal-ordered operators in \eqref{eq:Virasoro_commutations}, the Virasoro tensor does not act diagonally. 
Therefore the spectrum-generating currents are not (combinations of) quasi-primary fields, which hinders the computation of the conformal weights of the states on the Hilbert space.
In Section \ref{sec:large charge} a BMN-like limit which simplifies this issue will be presented.

%%%%%%%%%%%%%%%%%%%%%%%%%%%%%%%%%%%%%%%%%%%%%%%%%%%%%%%%%%
\section{Representations} \label{sec:representations}
%%%%%%%%%%%%%%%%%%%%%%%%%%%%%%%%%%%%%%%%%%%%%%%%%%%%%%%%%%

After having established that an extension of the affine Lie superalgebra $\mathfrak{g}_k \oplus \mathfrak{g}_{-k}$ (see \eqref{eq:mode_algebra} for the complete commutation relations) naturally acts on the Hilbert space of the theory, we go on and study possible representations of this algebra. 

There is immediately a severe problem arising, which hinders us from solving the complete theory. 
At the WZW-point, the representation theory of the algebra $\mathfrak{g}_k \oplus \mathfrak{g}_{-k}$ is very-well understood, see \cite{Gotz:2006qp} for the case of $\mathfrak{psu}(1,1|2)$. 
In particular the modes $Q^a_m$ define lowest weight representations on the Hilbert space,\footnote{This is of course not quite true for spectrally flowed representations, but for the sake of this argument, we restrict to the unflowed sector.} while the modes $\bar{Q}^{\bar{a}}_m$ define highest weight representations. 
Since the algebra depends only on $k$, this should not change when going away from the WZW-point. 
When adding the modes $P^a_m$ and $\bar{P}_m^{\bar{a}}$, it is natural to assume that they define the same kind of representations, since they form an affine algebra together with the modes $Q^a_m$ and $\bar{Q}^{\bar{a}}_m$. 
This however implies, by virtue of the commutation relations \eqref{eq:mode_algebra}, that the modes $A^{a\bar{a}}_m$ define neither highest nor lowest weight representations on the Hilbert space. 
This fact prevents us from computing conformal weights of excitations with both barred and unbarred oscillators. We will explain in the next section how to circumvent this problem in a BMN-like limit.

%%%%%%%%%%%%%%%%%%%%%%%%%%%%%%%
\subsection{Affine primaries}
%%%%%%%%%%%%%%%%%%%%%%%%%%%%%%%

Similarly to \cite{Ashok:2009xx, Benichou:2010rk} and analogously with the WZW-point, we define an affine primary state $\vert\Phi\rangle$ transforming in the representation $\mathcal{R}_0$ as follows:
\begin{align}\label{eq:affine primary definition QP}
\begin{aligned}
Q^a_m \ket{\Phi} &=0\, ,\ m>0\, ,&  Q^a_0 \ket{\Phi}&= t^a_{\mathcal{R}_0} \ket{\Phi}\, , & P^a_m \ket{\Phi} &=0\, ,\ m \ge 0 \\
\bar{Q}^{\bar{a}}_m \ket{\Phi} &=0\, ,\ m <0\, ,& \bar{Q}^{\bar{a}}_0 \ket{\Phi}&= t^{\bar{a}}_{\mathcal{R}_0} \ket{\Phi}\, ,& \bar{P}^{\bar{a}}_m,\ket{\Phi} &=0\, ,\ m \le 0 \, ,
\end{aligned}
\end{align}
where $t^a_{\mathcal{R}_0}$ are the generators of $\mathfrak{g}$ in the representation $\mathcal{R}_0$.
As we have mentioned, we cannot impose a highest or lowest weight condition on $A^{a\bar{a}}_m$. These conditions are consistent with the Jacobi identity.

One might also be worried with the fact that the anti-holomorphic Virasoro modes $\bar{L}_n$ can be expressed in terms of the unbarred oscillators, similarly to \eqref{eq:Virasoro_modes}. Our definition of affine primary states implies that the anti-holormophic Virasoro modes act in the opposite way than usual. Thus, it seems as if the spectrum is unbounded from below. However, due to the various identifications among the modes, several other states are removed from the spectrum. In particular negative energy states are consistently removed from the physical spectrum. 

We note in particular that the conformal weight of the ground state is now very easy to compute:
\begin{align}
L_0 \ket{\Phi}&=\frac{2f^2}{(1+k f^2)^2} (X^a X^a)_0 \ket{\Phi}=\frac{1}{2} f^2 \, t^a_{\mathcal{R}_0} \, t^a_{\mathcal{R}_0} \ket{\Phi}=\frac{1}{2} f^2 {\cal C}({\cal R}_0) \ket{\Phi}\, .
\end{align}
Thus, the conformal weight of an affine primary is given by
\be 
h(\ket{\Phi})=\tfrac{1}{2} f^2 \mathcal{C}(\mathcal{R}_0)\, , \label{eq:affine primary conformal weight}
\ee
where $\mathcal{C}(\mathcal{R}_0)$ denotes the quadratic Casimir of $\mathfrak{g}$ in $\mathcal{R}_0$. This matches with \cite{Benichou:2010rk, Bershadsky:1999hk}, and is the quantum analogue of \eqref{eq:ground_state_L0}.

%%%%%%%%%%%%%%%%%%%%%%%%%%%%%%%%%%%%%%%%%%%%%%%%%%%%%%%%%%
\subsection{Spectral flow} \label{subsec:spectral flow}
%%%%%%%%%%%%%%%%%%%%%%%%%%%%%%%%%%%%%%%%%%%%%%%%%%%%%%%%%%

We can also define so-called spectrally flowed representations of the mode algebra. 
For this, we introduce the following notation for Lie (super)algebras. Cartan-indices will be denoted by latin letters $i,j, \dots$, while roots will be denoted by greek letters $\alpha,\beta,\dots$. Hence $Q^i_0$ denote the Cartan-generators of $\mathfrak{g}$, while $Q^\alpha_0$ denote the step operators. 
We assume for ease of presentation that $\mathfrak{g}$ is simply-laced (this is in particular true for $\mathfrak{psu}(1,1|2))$, but the same analysis goes also through in the non simply-laced case. To match the usual conventions for $\mathfrak{su}(2)$ and $\mathfrak{psu}(1,1|2)$ (see Appendix~\ref{app:properties}), all roots are assumed to have length $1$. The commutation relations of the $Q^a_m$ with themselves take the following form in this basis \cite{DiFrancesco:1997nk}:
\begin{align}
[Q^i_m,Q^j_n]&= \kappa^{ij} k m\delta_{n+m,0}\, , \\
[Q^i_m,Q^\alpha_n]&=\alpha^i Q^\alpha_{m+n}\, , \\
[Q^\alpha_m,Q^\beta_n]&=\begin{cases}
k m \delta_{m+n,0}+ \kappa_{ij} \alpha^i Q^j_{m+n}\ , \quad & \alpha+\beta=0\, ,\\
{\cal N}_{\alpha,\beta} Q^{\alpha+\beta}_{m+n}\ ,  \quad & \alpha+\beta\text{ is a root}\, , \\
0\ , \quad &\text{otherwise}\, ,
\end{cases}
\end{align}
where $\kappa^{ij}=\tfrac{1}{2}\delta^{ij}$, and we used $\kappa^{\alpha\beta}=\delta_{\alpha+\beta,0}$.
Here ${\cal N}_{\alpha,\beta}$ are constants whose precise values do not play a r\^ole in the following.
Similarly, all other commutation relations of \eqref{eq:mode_algebra} can be written in this form. The action of the spectral flow on the modes is as follows:
\begin{align}
\begin{aligned} \label{eq:spectral flow}
\hat{Q}^i_m&=Q^i_m+\tfrac{1}{2} k w^i\delta_{m,0}\ , & \hat{P}^i_m&=P^i_m + \tfrac{1}{2}k w^i \delta_{m,0} - \tfrac{1}{2}k\kappa_{\bar{\imath}\bar{\jmath}} \bar{w}^{\bar{\imath}} A^{i\bar{\jmath}}_m\, , \\
\hat{Q}^\alpha_m&=Q^\alpha_{m+\alpha \cdot w/2}\ , & \hat{P}^\alpha_m&=P^\alpha_{m+\alpha \cdot w/2} - \tfrac{1}{2}k\kappa_{\bar{\imath}\bar{\jmath}} \bar{w}^{\bar{\imath}} A^{\alpha \bar{\jmath}}_{m+\alpha \cdot w/2}\, , \\
\hat{\bar{Q}}^{\bar{\imath}}_m&=\bar{Q}^{\bar{\imath}}_m - \tfrac{1}{2}k \bar{w}^{\bar{\imath}} \delta_{m,0}\ , & \hat{\bar{P}}^{\bar{\imath}}_m&=\bar{P}^{\bar{\imath}}_m - \tfrac{1}{2} k \bar{w}^{\bar{\imath}}\delta_{m,0} + \tfrac{1}{2}k\kappa_{ij} w^i A^{j\bar{\jmath}}_m\, , \\
\hat{\bar{Q}}^{\bar{\alpha}}_m&=\bar{Q}^{\bar{\alpha}}_{m + \bar{\alpha} \cdot \bar{w}/2}\ , & \hat{\bar{P}}^{\bar{\alpha}}_m&=\bar{P}^{\bar{\alpha}}_{m+\bar{\alpha} \cdot \bar{w}/2}+\tfrac{1}{2}k\kappa_{ij} w^i A^{j\bar{\alpha}}_{m+\bar{\alpha} \cdot \bar{w}/2}\, , \\
\hat{A}^{a\bar{a}}_m&=A^{a\bar{a}}_{m+a \cdot w/2+\bar{a} \cdot \bar{w}/2}\, ,
\end{aligned}
\end{align}
where $w^i$, $\bar{w}^{\bar{\imath}}$ are the spectral flow parameters.
Here, $\alpha \cdot w=\kappa_{ij} \alpha^i w^j$ is the inner product on the root space. We also used the notation $a \cdot w$, which equals $\alpha \cdot w$ if $a$ is a root index and zero if $a$ is a Cartan-index.
One can check that this indeed leaves the algebra \eqref{eq:mode_algebra} invariant. Note that the modes $A^{a\bar{a}}_m$ play a crucial r\^ole in defining this automorphism.  

\medskip

One can in particular investigate the effect of this automorphism on the energy-momentum tensor. For this, we observe that the spectral-flow symmetry in terms of the $X^a_m$ reads as follows:
\begin{multline}\label{eq:L spectral flow}
\hat{L}_n=L_n+\frac{1}{2}\kappa_{ij} w^i X^j_n + \frac{1}{2} \kappa_{\bar{\imath}\bar{\jmath}} \bar{w}^{\bar{\imath}} \bar{X}^{\bar{\jmath}}_n+\kappa_{ij}w^iw^j \frac{(1+k f^2)^2}{32f^2}\delta_{n,0}\\
+\kappa_{\bar{\imath}\bar{\jmath}} \bar{w}^{\bar{\imath}} \bar{w}^{\bar{\jmath}} \frac{(1-k f^2)}{32f^2} \delta_{n,0} - \kappa_{ij}\kappa_{\bar{\imath}\bar{\jmath}}w^i\bar{w}^{\bar{\imath}} \frac{1-k^2 f^4}{16f^2} A^{j\bar{\jmath}}_{n}\, .
\end{multline}
One may check that this indeed still satisfies the Virasoro algebra. 
The appearance of $X^j_n$ and $\bar{X}^{\bar{\jmath}}_n$ in a symmetric way is a very satisfying feature of this spectral flow symmetry. Unfortunately, also the modes $A^{j\bar{\jmath}}_n$ appear, which makes it generally hard to compute the effect of this spectral flow on states. 
Note also that these expressions reduce to the ones of \cite{Maldacena:2000hw, Gotz:2006qp} at the WZW-point.

\medskip

Similarly to the simplification in the representation theory, the spectral flow simplifies considerably when flowing only with the unbarred algebra, i.e.~$\bar{w}=0$. Then the field $A^{j\bar{\jmath}}_n$ disappears and the effect becomes computable. However, the physical spectrum seems to rather require $w=\bar{w}$ \cite{Gotz:2006qp}, so it is not clear whether it makes sense to look at states which are only partially spectrally flowed.
This deserves a better understanding.

%%%%%%%%%%%%%%%%%%%%%%%%%%%%%%%%%%%%%%%%%%%%%%%%%%%%%%%%%%
\section{The large charge limit} \label{sec:large charge}
%%%%%%%%%%%%%%%%%%%%%%%%%%%%%%%%%%%%%%%%%%%%%%%%%%%%%%%%%%

In this section we will consider a limit where all charges are sent to infinity. 
Since for affine algebras the charges are at most of the same order as their level $k$, we also require $k \to \infty$ at the same rate.
Finally, we require that $kf^2$ remains constant in the limit. 
In its applications to string theory (see Section \ref{sec:applications}), this will precisely correspond to the BMN-limit \cite{Berenstein:2002jq}. In this limit, the theory simplifies drastically, as we will see below.

%%%%%%%%%%%%%%%%%%%%%%%%%%%%%%%%%%%%%%%%%%%%%%%%%%%%%%%%%%
\subsection{The contraction of the mode algebra} \label{subsec:contraction}
%%%%%%%%%%%%%%%%%%%%%%%%%%%%%%%%%%%%%%%%%%%%%%%%%%%%%%%%%%

Let us consider the effect of this limit on the mode algebra \eqref{eq:mode_algebra}. 
The eigenvalues of the Cartan-generators $Q_0^i$ are of order $\mathcal{O}\big(k\big)$, since we assumed that all charges are of this order. 
The step-operators $Q_0^\alpha$ are of order $\mathcal{O}\big(k^{\frac{1}{2}}\big)$ (since their commutator gives back the Cartan-generators).
From \eqref{eq:affine primary definition QP} we know that $P^i_0\ket{\Phi}=0$, and so $P^i_0$ is not large even though it is a Cartan-generator. 
The modes $Q^a_m$ and $P^a_m$ for $m \ne 0$ are then of order $\mathcal{O}\big(k^{\frac{1}{2}}\big)$  in this limit, since their commutator gives the Cartan-generators and central terms. 
Hence, the Cartan-generators $Q_0^i$ are of order $\mathcal{O}\big(k\big)$, while all other oscillators are of order $\mathcal{O}\big(k^{\frac{1}{2}}\big)$. 
Keeping only the leading terms gives the following contraction of the mode algebra \eqref{eq:mode_algebra}
\begin{align}
\begin{aligned}
\,[Q^a_m,Q^b_n]&=\left(m k \kappa^{ab} +\mathrm{i}\tensor{f}{^{ab}_i} Q^i_0\right)\delta_{m+n,0}\, , \\
[Q^a_m,P^b_n]&=mk \kappa^{ab} \delta_{m+n,0}\, , \\
[P^a_m,P^b_n]&=0\, ,
\end{aligned}
\end{align}
and similarly for the barred oscillators. 
Furthermore note that the Cartan zero-modes $Q_0^i$ become central extensions of this almost-abelian algebra. 
Likewise, the field $A^{a\bar{a}}$ appears solely as a central extension. 
Since the $Q^i_0$ are central, we may replace them with their eigenvalues $\ell^i$ in the given representation.

Let us now look into the action of the modes $A_m^{a\bar{a}}$ in this limit.
In \cite{Benichou:2010rk} it was found that these modes are not all independent. In fact, we have the following relation:
\begin{equation}
m A^{a\bar{a}}_m =\frac{\mathrm{i}}{k}\tensor{f}{^a_{bc}} (P^c A^{b\bar{a}})_m=-\frac{\mathrm{i}}{k}\tensor{f}{^{\bar{a}}_{\bar{b}\bar{c}}} (\bar{P}^{\bar{c}} A^{a\bar{b}})_m\, ,
\end{equation}
From this relation we conclude that any non-zero mode is of order $\mathcal{O}\big(k^{-\frac{1}{2}}\big)$, whereas $A_0^{a\bar{a}}$ is of order $\mathcal{O}(1)$, see \eqref{eq: A definition}.
Therefore all non-zero modes are subleading in this limit.
Evaluating \eqref{eq: A definition} on the classical ground state \eqref{eq:classical ground state} yields $A^{i\bar{\imath}} = \kappa^{i\bar{\imath}}$.
It is then natural to assume
\be 
A_0^{\imath\bar{\imath}} \ket{\Phi}=\kappa^{\imath\bar{\imath}} \ket{\Phi}\, .
\ee
In particular, this is consistent with all commutation relations, as well as with all the identifications between the modes.

%%%%%%%%%%%%%%%%%%%%%%%%%%%%%%%%%%%%%%%%%%%%%%%%%%%%%%%%%%
\subsection{The spectrum-generating algebra}
%%%%%%%%%%%%%%%%%%%%%%%%%%%%%%%%%%%%%%%%%%%%%%%%%%%%%%%%%%

We now look at the commutation relations of $L_m$ with $Q^a_n$ and $P^a_n$, which follow from taking the appropriate limit of \eqref{eq:Virasoro_commutations}. Indeed, the commutator of $L_m$ with $Q^a_n$ does not change, while the commutator of $L_m$ with $P^a_n$ becomes
\begin{align}
[L_m,P^a_n]&=-k f^2 n Q^a_{m+n}-\tfrac{1}{2}(1-k f^2)n P^a_{m+n}-\mathrm{i} f^2 \tensor{f}{^a_{ic}} \ell^i P^c_{m+n} \nonumber\\
&=-k f^2 n  Q^a_{m+n}-\tfrac{1}{2}(1-k f^2)n P^a_{m+n}+f^2\kappa_{ij} a^j \ell^i P^a_{m+n}\, . \label{eq:Vir_limit}
\end{align}
In the second line, $a^j$ denotes $\alpha^j$ if $a$ is a root and $0$ if $a$ is a Cartan-index. 
Note that the coefficients of all three terms in \eqref{eq:Vir_limit} are of the same order $\mathcal{O}(1)$.
Thus we see that in this limit $L_m$ only mixes $Q^a_n$ and $P^a_n$, so we can simply find the eigenvectors. For $n \ne 0$, they are given by:
\be \label{eq: L0 eigenvectors BMN} 
J^a_{\pm,n}\equiv Q^a_n+\frac{-f^2(a\cdot \ell+kn)\pm\sqrt{n^2+2 (a \cdot \ell)  k f^4 n+ (a \cdot \ell)^2 f^4}}{2k f^2 n}P^a_n \, .
\ee
where $a \cdot \ell \equiv \kappa_{ij} a^j \ell^i$.
Their commutation relations with $L_m$ are given by
\be 
[L_m,J^a_{\pm, n}]=\frac{1}{2}\Big((a \cdot \ell)f^2-n\mp\sqrt{n^2+2(a \cdot \ell) k f^4 n+ (a \cdot \ell)^2 f^4} \Big)J^a_{\pm, {m+n}}\, .\label{eq:LJpm commutator}
\ee 
We can similarly diagonalize the barred modes and compute commutators with $\bar{L}_m$. The other relevant commutation relations are given by
\begin{align}
%[\bar{L}_m,J^a_{\pm, n}]=-\frac{1}{2}\Big(-(a \cdot \ell)f^2-n\pm\sqrt{n^2+2(a \cdot \ell) k f^4 n+ (a \cdot \ell)^2 f^4} \Big)J^a_{\pm, {m+n}}\, , \\
[L_m,\bar{J}^{\bar{a}}_{\pm, n}]=\frac{1}{2}\Big((\bar{a} \cdot \bar{\ell})f^2-n\mp\sqrt{n^2-2(\bar{a} \cdot \bar{\ell}) k f^4 n+ (\bar{a} \cdot \bar{\ell})^2 f^4} \Big)\bar{J}^{\bar{a}}_{\pm, {m+n}}\, , \label{eq:LJbar commutator}
%[\bar{L}_m,\bar{J}^{\bar{a}}_{\pm, n}]=-\frac{1}{2}\Big(-(\bar{a} \cdot \bar{\ell})f^2-n\pm\sqrt{n^2-2(\bar{a} \cdot \bar{\ell}) k f^4 n+ (\bar{a} \cdot \bar{\ell})^2 f^4} \Big)\bar{J}^{\bar{a}}_{\pm, {m+n}}\, . \label{eq:LbarJbarpm commutator}
\end{align}
Moreover, $[L_m-\bar{L}_m,J^a_{\pm,n}]=-n J^a_{\pm,n}$ and similarly for the barred oscillators. This was expected, since $L_m-\bar{L}_m$ measures the spin of the state which should be an integer.

\medskip

We thus seem to obtain four oscillators $J^a_{\pm,n}$, $\bar{J}^a_{\pm,n}$ generating the CFT spectrum in this limit, but in fact only two are independent.
Indeed, knowing the two currents $j_z$ and $j_{\bar{z}}$, for example, is enough to completely determine (up to an isometry transformation) the classical solution $g(z,\bar{z})$ using \eqref{eq:currrents def}.
The quantum version of this statement is the relation \cite{Benichou:2010rk}
\begin{align}
\begin{aligned}
\bar{Q}^a_m & = -\kappa_{ab}(Q^aA^{b\bar{a}})_m + \kappa_{ab}(P^aA^{b\bar{a}})_m \\
\bar{P}^a_m & = \kappa_{ab}(P^aA^{b\bar{a}})_m \, ,
\end{aligned}
\end{align}
between the modes introduced in \eqref{eq: Q and P defs}.
This relation allows us to express the actions of $\bar{Q}_m^a$, $\bar{P}_m^a$ on an affine primary, for example, in terms of the actions of $Q^a_m$ and $P^a_m$.\footnote{Note that the action of $A^{a\bar{a}}_m$ can be likewise, in principle, expressed in terms of the modes $Q^a_m$ and $P^a_m$.
%The precise way in which this is done will be explored elsewhere \cite{Eberhardt:2018}.
For our purposes it is enough to notice that the number of oscillators is reduced from four to two.}
More concretely, the semiclassical solutions suggests that we should identify the following two states up to a phase:
\begin{equation}\label{eq: states identified}
J^a_{\epsilon,n}\vert \ell^i,\bar{\ell}^i+\bar{a}^i\rangle \longleftrightarrow \bar{J}^{\bar{a}}_{\epsilon, n}\vert \ell^i + a^i,\bar{\ell}^i \rangle \, , 
\end{equation}
with $(a\cdot\ell)=-(\bar{a}\cdot\bar{\ell})$.
Here $\vert \ell^i,\bar{\ell}^i+\bar{a}^i\rangle$ and $\vert \ell^i + a^i,\bar{\ell}^i\rangle$ are affine primary states with charges $(\ell^i$, $\bar{\ell}^i+\bar{a}^i)$ and $(\ell^i + a^i$, $\bar{\ell}^i )$, respectively.
These two affine primary states can be obtained from each other by the action of the zero-modes.
It is easy to see that the charges of the states \eqref{eq: states identified} are the same, as well as their conformal dimension.

Thus, we can now generate the spectrum by considering solely the states
\be 
\ket{\Psi}\equiv \prod_{n=1}^\infty \prod_{i_n=1}^{N_n}J^{a_{i_n}}_{-,-n} \prod_{\bar{n}=1}^\infty \prod_{\bar{\imath}_{\bar{n}}=1}^{\bar{N}_{\bar{n}}}\bar{J}^{\bar{a}_{\bar{\imath}_{\bar{n}}}}_{-,\bar{n}} \ket{\Phi}\, .
\ee 
The conformal weight of these states follows from \eqref{eq:LJpm commutator} and \eqref{eq:LJbar commutator}:
\begin{multline}
h(\ket{\Psi})=h(\ket{\Phi})+\frac{1}{2}\sum_{n=1}^\infty \sum_{i_n=1}^{N_n} \Big((a_{i_n} \cdot \ell)f^2+n+\sqrt{n^2-2(a_{i_n} \cdot \ell) k f^4 n+ (a_{i_n} \cdot \ell)^2 f^4} \Big) \\
+\frac{1}{2}\sum_{\bar{n}=1}^\infty \sum_{\bar{\imath}_{\bar{n}}=1}^{\bar{N}_{\bar{n}}} \Big((\bar{a}_{\bar{\imath}_{\bar{n}}} \cdot \bar{\ell})f^2+\bar{n}+\sqrt{\bar{n}^2-2(\bar{a}_{\bar{\imath}_{\bar{n}}} \cdot \bar{\ell}) k f^4 \bar{n}+ (\bar{a}_{\bar{\imath}_{\bar{n}}} \cdot \bar{\ell})^2 f^4} \Big)\, .
\end{multline}
In later applications, we will always take the left- and right-moving representations of $\ket{\Phi}$ to coincide, i.e.~we will consider the diagonal modular invariant. Let us now restrict to this case, where $\ell=\bar{\ell}$. Then we use the notation $\bar{n}=-n$ for $n<0$ to write the formula in a compact way as follows:
\begin{multline}
h(\ket{\Psi})=\frac{1}{2} f^2 {\cal C}({\cal R}_0)\\
+\frac{1}{2}\sum_{\genfrac{}{}{0pt}{}{n=-\infty}{n \ne 0}}^\infty \sum_{i_n=1}^{N_n} \Big((a_{i_n} \cdot \ell)f^2+n+\sqrt{n^2-2(a_{i_n} \cdot \ell) k f^4 |n|+ (a_{i_n} \cdot \ell)^2 f^4} \Big)\, , \label{eq:BMN conformal weight}
\end{multline}
where we also inserted the conformal weight of the affine primary \eqref{eq:affine primary conformal weight}. This is the main result of this section.

The spin $s=h-\bar{h}$ of the state is given by
\be 
s(\ket{\Psi})=h(\ket{\Psi})-\bar{h}(\ket{\Psi})=\sum_{n=-\infty}^\infty n N_n\, .
\ee
In particular, it is integer, which is a consistency check of our analysis. One may also check that this formula reduces to the correct conformal weight at the WZW-point $kf^2=1$.

%%%%%%%%%%%%%%%%%%%%%%%%%%%%%%%%%%%%%%%%%%%%%%%%%%%%%%%%%%
\subsection{Characters}
%%%%%%%%%%%%%%%%%%%%%%%%%%%%%%%%%%%%%%%%%%%%%%%%%%%%%%%%%%

We can work out the characters of the representation we just found. Since $k$ is large, the Verma-module does not contain any null-vectors. We can directly read from \eqref{eq:BMN conformal weight} the character of such a representation:
\begin{align}
\chi(\tau,\bar{\tau})&=|\chi_0|^2\prod_{\genfrac{}{}{0pt}{}{n=-\infty}{n \ne 0}}^\infty  \prod_a \Big(1-\abs{q}^{(a \cdot \ell)f^2+\sqrt{n^2- 2(a \cdot \ell) k f^4 |n|+ (a  \cdot \ell)^2 f^4}} q^{\frac{n}{2}} \bar{q}^{-\frac{n}{2}} \Big)^{-|a|}\, .
\end{align}
Here $a$ runs over the complete Lie superalgebra and $|a|=1$ if the index is bosonic and $|a|=-1$ if it is fermionic. Also, $\chi_0$ denotes the character of the zero-mode algebra of the representation ${\cal R}_0$. We have not included chemical potentials in the formula, their inclusion is straightforward.

%%%%%%%%%%%%%%%%%%%%%%%%%%%%%%%%%%%%%%%%%%%%%%%%%%%%%%%%%%
\section{Applications to string theory} \label{sec:applications}
%%%%%%%%%%%%%%%%%%%%%%%%%%%%%%%%%%%%%%%%%%%%%%%%%%%%%%%%%%

In this section, we will apply the formalism we constructed in the previous sections to string theory on the backgrounds $\mathrm{AdS}_3 \times \mathrm{S}^3 \times \mathbb{T}^4$ and $\mathrm{AdS}_3 \times \mathrm{S}^3 \times {\rm S}^3 \times {\rm S}^1$.\footnote{A similar treatment applies to $\mathrm{AdS}_3 \times \mathrm{S}^3 \times \mathrm{K3}$.} For this, our starting point is the hybrid formalism for $\mathrm{AdS}_3 \times \mathrm{S}^3 \times \mathbb{T}^4$ \cite{Berkovits:1999im}, in which the sigma-model on the supergroup $\mathrm{PSU}(1,1|2)$ features prominently.

%%%%%%%%%%%%%%%%%%%%%%%%%%%%%%%%%%%%%%%%%%%%%%%%%%%%%%%%%%
\subsection{Review of the hybrid formalism}
%%%%%%%%%%%%%%%%%%%%%%%%%%%%%%%%%%%%%%%%%%%%%%%%%%%%%%%%%%

Since the hybrid formalism is rather involved, we will only review the most important features for the case $\mathrm{AdS}_3 \times \mathrm{S}^3 \times \mathbb{T}^4$. 
We need the following three ingredients:
\begin{enumerate}
\item A sigma-model on the supergroup $\mathrm{PSU}(1,1|2)$, eq.~\eqref{eq:action}.
\item A topologically twisted $c=6$ $\mathcal{N}=4$ CFT. In our case, this is the topologically twisted CFT on $\mathbb{T}^4$.
\item Two additional ghost fields $\rho$ and $\sigma$. They couple in general to the other fields in the theory.
\end{enumerate}
From the fields of the theory, one can define a twisted $\mathcal{N}=4$ superconformal algebra. 
Physical states are identified with the double cohomology of this twisted $\mathcal{N}=4$ superconformal algebra with unit R-charge. 

As discussed in Section \ref{sec:current}, the sigma-model on the supergroup $\mathrm{PSU}(1,1|2)$ is a CFT and possesses two parameters $k$ and $f^2$. As seen in the previous sections, $k$ appears as the level of an $\mathfrak{su}(2)$-current algebra and is hence quantised as usual in WZW models. In string theory, this is interpreted as the number of NS5-branes creating the background geometry, and therefore also their total charge $Q_5^{\text{NS}}\equiv k$. On the other hand, $f^{-1}$ describes the radius of $\mathrm{AdS}_3$ and $\mathrm{S}^3$, which we denote by $R_{\text{AdS}}$. The relation with the D5-brane charge $Q_5^{\text{RR}}$ can be found for large radii using supergravity, and reads \cite{Berkovits:1999im, Maldacena:1998bw, David:2002wn, Bershadsky:1999hk}:
\be 
\frac{1}{f^2}=\frac{R_{\text{AdS}}^2}{\alpha'}=\sqrt{\big(Q_5^{\mathrm{NS}}\big)^2+g^2\big(Q_5^{\mathrm{RR}}\big)^2}\, . \label{eq:relation f and Qs}
\ee
Here $g$ is the ten-dimensional string coupling constant. As already mentioned in the introduction, this explains why $f^{-2}$ is not quantised in the worldsheet description: Since we are treating the string perturbatively, $g$ is small and hence $Q_5^{\mathrm{RR}}$ has to be of order $g^{-1}$ to have a visible effect on $f^{-2}$. Thus, it is effectively continuous in the worldsheet theory. In a full non-perturbative description of string theory, also $f^{-2}$ would become quantised. Note that \eqref{eq:relation f and Qs} gives the physical reason for the parameter range
\be 
-1 \le kf^2 \le 1\, .
\ee
Negative values of $kf^2$ correspond to anti-branes, we will not consider them in the following. %We will see below that the same bound is imposed by requiring the sigma-model to be unitary.

The FS1- and D1-brane charges enter as follows in the hybrid formalism. Supersymmetry imposes that the ratios $Q_5/Q_1$ agree for NS- and R-R-fields:
\be 
\frac{Q_5^{\mathrm{NS}}}{Q_1^{\mathrm{NS}}}=\frac{Q_5^{\mathrm{RR}}}{Q_1^{\mathrm{RR}}}\, .
\ee
Finally, $Q_1^{\mathrm{RR}}$ determines the volume of the compactification manifold $\mathbb{T}^4$ as
\be 
v=f^2 g Q_1^{\mathrm{RR}}\, ,
\ee
but it does not enter directly in the $\mathrm{PSU}(1,1|2)$-sigma model.

%%%%%%%%%%%%%%%%%%%%%%%%%%%%%%%%%%%%%%%%%%%%%%%%%%%%%%%%%%
\subsection{The BMN limit}
%%%%%%%%%%%%%%%%%%%%%%%%%%%%%%%%%%%%%%%%%%%%%%%%%%%%%%%%%%

The plane-wave or Berenstein-Maldacena-Nastase (BMN) limit \cite{Berenstein:2002jq} is the following limiting case of the theory:
\be 
j,\, \ell,\, k,\, f^{-2} \to \infty\, ,
\ee
with all their ratios remaining constant in the limit. Here, $j$ and $\ell$ are the eigenvalues of the Cartan-generators of $\mathfrak{sl}(2,\mathds{R})$ and $\mathfrak{su}(2)$, respectively. Also, the BMN limit is near-BPS, meaning that $j-\ell$ is kept finite in the limit.

\medskip
The complete action for the worldsheet theory reads
\be 
\mathcal{S}=f^{-2} \big(\mathcal{S}_0+\mathcal{S}_1)+k \mathcal{S}_\mathrm{WZ}+\mathcal{S}_\text{ghost}\, ,
\ee
where $f^{-2}\mathcal{S}_0+k \mathcal{S}_\mathrm{WZ}$ is the action of the $\mathrm{PSU}(1,1|2)$-sigma model as in eq.~\eqref{eq:action}. 
Furthermore, $\mathcal{S}_1$ are ghost couplings \cite[eq.~(8.39)]{Berkovits:1999im}. 
These ghost couplings are bilinear in the fermionic currents, and the ghosts which appear are at worst of order $\mathcal{O}(1)$. 
As discussed in Subsection~\ref{subsec:contraction}, the fermionic currents scale in the BMN limit as $\mathcal{O}\big(k^{\frac{1}{2}}\big)$. 
Therefore $\mathcal{S}_1$ is of order $\mathcal{O}\big(k)$. 
On the other hand, $\mathcal{S}_0$ is bilinear in the bosonic currents, which can be of order $\mathcal{O}(k)$.
Hence $\mathcal{S}_0$ is of order $\mathcal{O}(k^2)$. 
Thus, the ghost couplings are very much suppressed in the BMN limit and can be neglected. 
This was to be expected, since the ghost couplings vanish in flat space and the BMN limit is an almost-flat space approximation.

\medskip
Berenstein, Maldacena and Nastase derived in \cite{Berenstein:2002jq} a formula for the string spectrum in this limit:
\be 
\Delta-L=\sum_{n=-\infty}^\infty N_n \sqrt{1\pm\frac{2nk}{L}+\frac{n^2}{L^2f^4}}+\frac{1}{L f^2} \big(L_0^{\mathbb{T}^4}+\bar{L}_0^{\mathbb{T}^4}\big)+\mathcal{O}\big(k^{-1}\big)\, . \label{eq:BMN formula}
\ee
Here, $L_0^{\mathbb{T}^4}$ is the conformal weight coming from the torus excitations, $L=\ell+\bar{\ell}$ is the total $\mathfrak{su}(2)$-spin from both left- and right-movers, and $\Delta=j+\bar{\jmath}$ is the scaling dimension of the dual CFT. 
Since $\Delta$ and $L$ are both large, but their difference is finite, this is a near-BPS limit. The summation goes over the different worldsheet oscillators, where $n<0$ refers to right-movers and $n>0$ to left-movers. Also, $N_n$ is the occupation number of the respective mode. Level-matching translates into
\be 
\sum_{n=-\infty}^{\infty} n N_n= \bar{L}_0^{\mathbb{T}^4}-L_0^{\mathbb{T}^4}\, ,\label{eq:level matching}
\ee
in this language. Notice that while the RHS of \eqref{eq:BMN formula} contains $L$, we could have also written $\Delta$ since these quantities differ only by subleading terms. 
We also assume that only finitely many occupation numbers $N_n$ are non-zero.

%%%%%%%%%%%%%%%%%%%%%%%%%%%%%%%%%%%%%%%%%%%%%%%%%%%%%%%%%%
\subsection{Reproducing the BMN formula of AdS$_{\text 3} \boldsymbol\times $S$^{\text 3} \boldsymbol\times\mathbb{T}^{\text{4}}$ from the worldsheet}
%%%%%%%%%%%%%%%%%%%%%%%%%%%%%%%%%%%%%%%%%%%%%%%%%%%%%%%%%%

We are finally in the position to reproduce \eqref{eq:BMN formula} from the worldsheet. 
For this, we note that the BMN limit coincides on the worldsheet precisely with the large charge limit we considered in Section~\ref{sec:large charge}. 
So we may start with \eqref{eq:BMN conformal weight}, which we derived in the last section. Requiring the state to be level-matched, i.e.~\eqref{eq:level matching} to be satisfied, \eqref{eq:BMN conformal weight} simplifies to
\begin{align}
h(\ket{\Psi})&=
\frac{1}{2} f^2 {\cal C}({\cal R}_0)+L_0^{\mathbb{T}^4}\nonumber\\
&\quad+\frac{1}{2}\sum_{\genfrac{}{}{0pt}{}{n=-\infty}{n \ne 0}}^\infty \sum_{i_n=1}^{N_n} \Big((a_{i_n}\!\! \cdot \ell)f^2+n+\sqrt{n^2- 2(a_{i_n}\!\! \cdot \ell) k f^4 |n|+ (a_{i_n}\!\! \cdot \ell)^2 f^4} \Big)\,, \nonumber\\
&=\frac{1}{2} f^2 {\cal C}({\cal R}_0)+\frac{1}{2}\big(L_0^{\mathbb{T}^4}+\bar{L}_0^{\mathbb{T}^4}\big)\nonumber\\
&\quad+\frac{1}{2}\sum_{\genfrac{}{}{0pt}{}{n=-\infty}{n \ne 0}}^\infty \sum_{i_n=1}^{N_n} \Big((a_{i_n} \!\!\cdot \ell)f^2+\sqrt{n^2- 2(a_{i_n}\!\! \cdot \ell) k f^4 |n|+ (a_{i_n}\!\! \cdot \ell)^2 f^4} \Big)\, . \label{eq:BMN conformal weight level matched}
\end{align}
We included a possible conformal weight from the torus. 
For the case of $\mathfrak{psu}(1,1|2)$, the Casimir equals \eqref{eq:psu112 casimir}:
\be 
{\cal C}({\cal R}_0)=-2j_0(j_0-1)+2\ell_0(\ell_0+1)\, , \label{eq:ground state casimir}
\ee
where $j_0$ and $\ell_0$ denote the $\mathfrak{sl}(2,\mathds{R})$-spin and the $\mathfrak{su}(2)$-spin of the ground state $\ket{\Phi}$, respectively. 
Note that the generic charges $\ell^i$ of Section \ref{sec:large charge} correspond now to $j_0$ and $\ell_0$. 
Furthermore, notice that solving the mass-shell condition $h(\ket{\Psi})=0$ will imply that $j_0=\ell_0+\mathcal{O}(1)$. 
To this order, we may therefore replace $j_0$ everywhere by $\ell_0$, except in \eqref{eq:ground state casimir}. 
Thus we have $a \cdot \ell=a \ell_0$, where $a$ takes the following values for the generators of $\mathfrak{psu}(1,1|2)$:
\begin{align}
J^3:&\  0\, , & J^\pm:&\ \mp 2\, , & K^3:&\ 0\, , &K^\pm:&\ \pm 2\, , &S^{\pm\pm\alpha}:&\ 0\, , &S^{\pm\mp\alpha}:&\ \mp 2\, .
\end{align}
For the complete commutation relations of the affine algebra $\mathfrak{psu}(1,1|2)_k$ in this basis, see Appendix~\ref{subapp:psu112}. 
With all this in mind, we now solve \eqref{eq:BMN conformal weight level matched} for $j_0$ and expand the result in orders of the characteristic scale $k$ to obtain:
\begin{multline}\label{eq:j0 solved}
j_0 =\ell_0+1+\frac{1}{4}\sum_{\genfrac{}{}{0pt}{}{n=-\infty}{n \ne 0}}^\infty \sum_{i_n=1}^{N_n} \Bigg(a_{i_n}+\sqrt{a_{i_n}^2- \frac{2 a_{i_n} k |n|}{\ell_0}+\frac{n^2}{\ell_0^2f^4}} \Bigg)\\
+\frac{1}{4\ell_0 f^2} \big(L_0^{\mathbb{T}^4}+\bar{L}_0^{\mathbb{T}^4}\big)+\mathcal{O}\big(k^{-1}\big)\, . 
\end{multline}
The summand $1$ comes from the fact that $j_0$ and $\ell_0$ measure the $\mathfrak{sl}(2,\mathds{R})$ and $\mathfrak{su}(2)$-spin of the highest weight state. As one can see from the structure of a typical multiplet of $\mathfrak{psu}(1,1|2)$ (see \ref{eq:psu112 multiplet}), the state with the lowest $j-\ell$ is not the highest weight state and it has precisely $j-\ell$ lowered by one.\footnote{In fact there are four such states.}

Finally, we have to take into account the contribution of the oscillators to the $\mathfrak{sl}(2,\mathds{R})$ and $\mathfrak{su}(2)$-spins. We notice that $a$ measures precisely the difference of $\mathfrak{sl}(2,\mathds{R})$-spin with $\mathfrak{su}(2)$-spin of every oscillator:
\be 
a=2 \times (\text{$\mathfrak{su}(2)$-spin})-2 \times (\text{$\mathfrak{sl}(2,\mathds{R})$-spin})\, .
\ee
Combining these observations, we find
\be 
j-\ell=j_0-\ell_0-1-\frac{1}{2}\sum_{n=1}^\infty \sum_{i_n=1}^{N_n} a_{i_n}\ , \quad \bar{\jmath}-\bar{\ell}=j_0-\ell_0-1-\frac{1}{2}\sum_{n=-\infty}^{-1} \sum_{i_n=1}^{N_n} a_{i_n}\, ,
\ee
where $j$, $\ell$ denote the $\mathfrak{sl}(2,\mathds{R})$ and $\mathfrak{su}(2)$ spins of the state $\vert \Psi\rangle$, respectively.
Defining $\Delta=j+\bar{\jmath}$ and $L=\ell+\bar{\ell}$, and combining all the ingredients, we finally obtain
\begin{align}
\Delta-L&=2j_0-2\ell_0-2-\frac{1}{2}\sum_{\genfrac{}{}{0pt}{}{n=-\infty}{n \ne 0}}^\infty \sum_{i_n=1}^{N_n} a_{i_n} \nonumber \\
&=\sum_{\genfrac{}{}{0pt}{}{n=-\infty}{n \ne 0}}^\infty \sum_{i_n=1}^{N_n} \sqrt{\frac{a_{i_n}^2}{4}- \frac{a_{i_n} k |n|}{2\ell_0}+\frac{n^2}{4\ell_0^2f^4}}+\frac{1}{2\ell_0 f^2} \big(L_0^{\mathbb{T}^4}+\bar{L}_0^{\mathbb{T}^4}\big) +\mathcal{O}\big(k^{-1}\big)\nonumber \\
&=\sum_{\genfrac{}{}{0pt}{}{n=-\infty}{n \ne 0}}^\infty \sum_{i_n=1}^{N_n} \sqrt{\frac{a_{i_n}^2}{4}- \frac{a_{i_n} k |n|}{L}+\frac{n^2}{L^2f^4}} +\frac{1}{L f^2} \big(L_0^{\mathbb{T}^4}+\bar{L}_0^{\mathbb{T}^4}\big)+\mathcal{O}\big(k^{-1}\big)\, .\label{eq:BMN formula from worldsheet}
\end{align}
Finally, we impose the remaining physical state conditions which have the effect of removing all oscillators with $a_{i_n}=0$.
This can be most easily seen by comparing the pure NS-NS case with the partition function derived in \cite{Maldacena:2000kv,Raju:2007uj}.
The cohomological argument was given in \cite{Gerigk:2012cq}.
Therefore the physical oscillators are the ones with $a_{i_n}=\pm 2$, and thus the physical spectrum reads
%To make contact with \eqref{eq:BMN formula} we choose the set of oscillators chosen in its derivation (see the paragraph following \eqref{eq:BMN formula}), and for which $a_{i_n}=-2$ for $n>0$ and $a_{i_n}=2$ for $n<0$.
%We obtain
\be 
\Delta-L=\sum_{\genfrac{}{}{0pt}{}{n=-\infty}{n \ne 0}}^\infty N_n \sqrt{1\pm\frac{2k n}{L}+\frac{n^2}{L^2f^4}}+\frac{1}{L f^2} \big(L_0^{\mathbb{T}^4}+\bar{L}_0^{\mathbb{T}^4}\big) +\mathcal{O}\big(k^{-1}\big)\, ,
\ee
which matches the BMN formula \eqref{eq:BMN formula}.
This concludes the derivation of the BMN formula from the worldsheet. 

\medskip

One can furthermore confirm that the analysis holds also true in the spectrally flowed sectors. For this, we choose $w \equiv w^{\mathfrak{sl}(2,\mathds{R})}=-\bar{w}^{\mathfrak{sl}(2,\mathds{R})}=w^{\mathfrak{su}(2)}=-\bar{w}^{\mathfrak{su}(2)}$.\footnote{The sign for the barred spectral flow parameters seem peculiar, but this is again related to our mode conventions for the barred modes.} We spectrally flow the ground state on top of which we build the spectrum, e.g.~the state in the $\mathfrak{psu}(1,1|2)$-multiplet with quantum numbers $(j_0-1,\ell_0)$. From \eqref{eq:L spectral flow}, we conclude that on this state
\be 
\hat{L}_0 \ket{\Phi}= f^2 (-\hat{j}_0(\hat{j}_0-1)+\hat{\ell}_0(\hat{\ell}_0+1)) \ket{\Phi}=\tfrac{1}{2}f^2 \mathcal{C}\big(\hat{\mathcal{R}}_0 \big) \ket{\Phi}\ ,
\ee
where $\hat{j}_0=j_0+\tfrac{kw}{2}$ and $\hat{\ell}_0=\ell_0+\tfrac{kw}{2}$ are the spectrally flowed spins of the ground state, see eq.~\eqref{eq:spectral flow}. After this, we can apply hatted oscillators on this spectrally flowed ground state to generate a state in this new representation. Since the spectral flow is an automorphism of the spectrum-generating algebra, the derivation is from hereon exactly the same as before, except that everything is replaced by spectrally flowed quantities. We obtain precisely \eqref{eq:BMN formula from worldsheet}, except that all quantities are now spectrally flowed. So we conclude that \eqref{eq:BMN formula from worldsheet} continues to hold true in the spectrally flowed sectors.

%%%%%%%%%%%%%%%%%%%%%%%%%%%%%%%%%%%%%%%%%%%%%%%%%%%%%%%%%%
\subsection{The case of AdS$_{\text 3} \boldsymbol\times $S$^{\text 3} \boldsymbol\times$S$^{\text 3} \boldsymbol\times$S$^{\text 1}$} \label{subsec:AdS3S3S3S1}
%%%%%%%%%%%%%%%%%%%%%%%%%%%%%%%%%%%%%%%%%%%%%%%%%%%%%%%%%%

We can similarly treat the background $\mathrm{AdS}_3 \times \mathrm{S}^3 \times \mathrm{S}^3 \times \mathrm{S}^1$, which has recently attracted considerable attention \cite{Gukov:2004ym, Tong:2014yna, Eberhardt:2017fsi, Eberhardt:2017pty, Gaberdiel:2018rqv, Baggio:2017kza, Dei:2018yth}.

Currently there exists no hybrid formalism \`a la Berkovits, Vafa and Witten for this background.
Nevertheless we expect that in the BMN limit, in analogy with $\mathrm{AdS}_3 \times \mathrm{S}^3 \times \mathbb{T}^4$, the theory can be described by a sigma model on $\mathrm{D}(2,1;\alpha)$ together with the theory on $\mathrm{S}^1$ and free ghosts. 
The bosonic part of the Lie supergroup $\mathrm{D}(2,1;\alpha)$ is $\mathrm{AdS}_3 \times \mathrm{S}^3 \times \mathrm{S}^3$, with the parameter $\alpha$ giving the ratio of the radii of the two spheres (for more details, see Appendix~\ref{subapp:d21alpha}). 
Representations are now labelled by the three spins $(j_0,\ell_0^+,\ell_0^-)$. 
The Casimir of such a representation is given by
\be 
{\cal C}(j_0,\ell_0^+,\ell_0^-)=-2j_0(j_0-1)+2\cos^2 \varphi\ \ell_0^+(\ell_0^++1)+2\sin^2 \varphi\ \ell_0^-(\ell^-+1)\, ,
\ee
where we introduced the angle $0 \le \varphi \le \tfrac{\pi}{2}$ such that $\alpha=\cot^2 \varphi$.\footnote{We thank Andrea Dei for bringing this parametrization to our attention.} In the limit we are taking, we set
\be 
\ell_0^+=\frac{\cos \omega}{\cos \varphi} \ell_0\,,\quad \ell_0^-=\frac{\sin \omega}{\sin \varphi} \ell_0\, ,
\ee
where $0 \le \omega\le \frac{\pi}{2}$ is another angle parametrizing the ratio of the spins as these are taken to infinity. 
Then as before we have $j_0=\ell_0+{\cal O}(1)$, and hence again $a\cdot \ell$ can be replaced with $a\ell_0$, where $a$ takes the following values for the different elements of the superalgebra $\mathfrak{d}(2,1;\alpha)$:
\begin{align}
\begin{aligned}
K^{(+)\pm}:&\ \pm 2\cos(\varphi)\cos(\omega)\, , &K^{(-)\pm}:&\ \pm 2\sin(\varphi)\sin(\omega)\, , & J^{\pm}: & \ \mp 2\, , \\
S^{\pm\pm\pm}:&\ \mp 2\sin^2 \Big(\frac{\varphi-\omega}{2}\Big)\, , &S^{\pm\mp\pm}:&\ \mp 2\cos^2\Big(\frac{\varphi+\omega}{2}\Big)\, , \\
S^{\pm\pm\mp}:&\ \mp 2\sin^2 \Big(\frac{\varphi+\omega}{2}\Big)\, , 
&S^{\pm\mp\mp}:&\ \mp 2\cos^2\Big(\frac{\varphi-\omega}{2}\Big)\, .
\end{aligned}
\end{align}
All three Cartan-generators $J^3$ and $K^{(\pm)3}$ have $a=0$. Solving the mass-shell condition for $j_0$ as in \eqref{eq:j0 solved} yields 
\begin{multline}
j_0=\ell_0+\cos^2\Big(\frac{\varphi-\omega}{2}\Big)+\frac{1}{4}\sum_{\genfrac{}{}{0pt}{}{n=-\infty}{n \ne 0}}^\infty \sum_{i_n=1}^{N_n} \Bigg(a_{i_n}+\sqrt{a_{i_n}^2- \frac{2 a_{i_n} k |n|}{\ell_0}+\frac{n^2}{\ell_0^2f^4}} \Bigg)\\
+\frac{1}{4\ell_0 f^2} \big(L_0^{\mathrm{S}^1}+\bar{L}_0^{\mathrm{S}^1}\big)+\mathcal{O}\big(k^{-1}\big)\, ,
\end{multline}
where we again used the mass-shell condition to simplify the result. 
We recognize that in this case, $a$ measures
\begin{multline} 
a=2 \cos(\varphi)\cos(\omega)\times (\text{$\mathfrak{su}(2)^+$-spin})\\+2 \sin(\varphi)\sin(\omega)\times (\text{$\mathfrak{su}(2)^-$-spin})-2 \times (\text{$\mathfrak{sl}(2,\mathds{R})$-spin})\, .
\end{multline}
Defining $\ell=\cos (\varphi) \cos (\omega)\ell^++\sin (\varphi) \sin (\omega)\ell^-$ and $L=\ell+\bar{\ell}$, $\Delta=j+\bar{\jmath}$, the same steps as before yield the final result
\begin{multline}
\Delta-L
=-\sin^2\Big(\frac{\varphi-\omega}{2}\Big)+\sum_{\genfrac{}{}{0pt}{}{n=-\infty}{n \ne 0}}^\infty \sum_{i_n=1}^{N_n} \sqrt{\frac{a_{i_n}^2}{4}- \frac{a_{i_n} k |n|}{L}+\frac{n^2}{L^2f^4}}\\+ \frac{1}{L f^2} \big(L_0^{\mathrm{S}^1}+\bar{L}_0^{\mathrm{S}^1}\big)+\mathcal{O}\big(k^{-1}\big)\, ,\label{eq:BMN formula from worldsheet AdS3S3S3S1}
\end{multline}
which formally coincides with \eqref{eq:BMN formula from worldsheet}, except for the constant squared sine. 
As before, in order to choose the state in a typical $\mathfrak{d}(2,1;\alpha)$ multiplet with smallest $\Delta-L$, a constant term 1 was included in the relations between $j-\ell$ and $j_0-\ell_0$.

Notice that the BPS condition for $\mathfrak{d}(2,1;\alpha)$ takes the form \eqref{eq: d21alpha BPS}
\be 
\Delta_{\mathrm{BPS}}=\cos^2(\varphi) L^++\sin^2(\varphi)L^-=\cos(\varphi-\omega)L\, ,
\ee
and so
\begin{multline} 
\Delta-\Delta_{\mathrm{BPS}}=(2L-1)\sin^2\Big(\frac{\varphi-\omega}{2}\Big)+\sum_{\genfrac{}{}{0pt}{}{n=-\infty}{n \ne 0}}^\infty \sum_{i_n=1}^{N_n} \sqrt{\frac{a_{i_n}^2}{4}- \frac{a_{i_n} k |n|}{L}+\frac{n^2}{L^2f^4}}\\+ \frac{1}{L f^2} \big(L_0^{\mathrm{S}^1}+\bar{L}_0^{\mathrm{S}^1}\big)+\mathcal{O}\big(k^{-1}\big)\, .
\end{multline}
In particular, since all terms on the right-hand side are positive,\footnote{Since $L \gg 1$, $2L-1$ is also positive.} we see that this is only a near-BPS expansion if $\varphi=\omega$, i.e.~$L^+=L^-$. Hence, all BPS states on the background $\mathrm{AdS}_3 \times \mathrm{S}^3 \times \mathrm{S}^3 \times \mathrm{S}^1$ have (in the large $k$ limit) $\ell^+=\ell^-$. This was recently shown in \cite{Eberhardt:2017fsi, Eberhardt:2017pty, Baggio:2017kza}, our calculation confirms the result again.
The squared sine of \eqref{eq:BMN formula from worldsheet AdS3S3S3S1} then vanishes for near-BPS states.

%%%%%%%%%%%%%%%%%%%%%%%%%%%%%%%%%%%%%%%%%%%%%%%%%%%%%%%%%%%%%%%%%%%%%%%
\section{Discussion} \label{sec:discussion}
%%%%%%%%%%%%%%%%%%%%%%%%%%%%%%%%%%%%%%%%%%%%%%%%%%%%%%%%%%%%%%%%%%%%%%%
In this paper we considered string theory on $\mathrm{AdS}_3$-backgrounds by employing the hybrid formalism of Berkovits, Vafa and Witten \cite{Berkovits:1999im}. 
This lead us to a study of sigma-models on Lie supergroups. 
We first found exact classical solutions of the model, which were suggestive of the BMN-formula when interpreted semiclassically. 
Starting from Section~\ref{sec:current}, we turned to a systematic study of the quantum mechanical sigma-model. 
In a BMN-like limit, where all charges become large and only finitely many excitations are considered, a complete solution to the model was found. 
We subsequently applied our findings to string theory on the backgrounds ${\rm AdS}_3 \times {\rm S}^3 \times {\cal M}_4$ with ${\cal M}_4=\mathbb{T}^4$, ${\rm K3}$ and ${\rm S}^3 \times {\rm S}^1$ via the hybrid formalism, where it allowed us to derive the complete plane-wave spectrum.

Our results provide a direct link between Green-Schwarz-like computations and worldsheet methods to determine the spectrum. 
The tools developed in this paper seem much more powerful than necessary to derive the plane-wave spectrum.
We can in principle derive the exact conformal weights of arbitrary one-sided excitations (i.e.~constructed using only unbarred modes), at least level by level. 
With the help of this, we can confirm some well-known conjectures explicitly, such as the fact that long strings disappear from the string spectrum away from the WZW-point. 
We can also retrieve the missing chiral primaries in the spacetime BPS spectrum \cite{Seiberg:1999xz, Argurio:2000tb, Raju:2007uj}.
We will report on this elsewhere \cite{Eberhardt:2018}.

We have presented in Section~\ref{sec:classical} an exact one-excitation solution of the classical theory. 
One can hope to extend this result to multi-particle excitations by employing integrability methods \cite{Gromov:2007aq,Beisert:2010jr,Vicedo:2011zz}.
In particular, the presented solution corresponds to a one-cut solution of the spectral curve. 
In principle, the spectral curve can be used to extend the result to multi-cut solutions. This would provide a way to compute the spectrum of string theory beyond the plane-wave limit. 

We expect that the analysis can be extended to other backgrounds like $\mathrm{AdS}_5 \times \mathrm{S}^5$, $\mathrm{AdS}_4 \times \mathbb{CP}^3$ and $\mathrm{AdS}_2 \times \mathrm{S}^2 \times \mathbb{T}^6$, where similar supergroup actions exist \cite{Metsaev:1998it, Arutyunov:2008if, Stefanski:2008ik, Zhou:1999sm, Berkovits:1999zq}. They feature the supergroups $\mathrm{PSU}(2,2|4)$, $\mathrm{OSP}(6|2,2)$ and $\mathrm{PSU}(1,1|2)$, which all have vanishing dual Coxeter numbers. However, the backgrounds require us to consider cosets of these supergroups, so one should effectively consider a coset of the current algebra considered in this paper. 
We expect that this can be worked out, but have not tried to do so.

We also hope that these considerations can be used to shed light on the emergence of a higher spin symmetry in the dual CFT. The backgrounds ${\rm AdS}_3 \times {\rm S}^3 \times \mathcal{M}_4$ are conjectured to lie on the same moduli space as the symmetric product orbifold of $\mathcal{M}_4$ \cite{Maldacena:1997re, Dijkgraaf:1998gf, Eberhardt:2017pty}.\footnote{In the case of $\mathcal{M}_4={\rm S}^3 \times {\rm S}^1$, this is only true for an integer ratio of the two D5-brane charges.} 
However, it is not clear where the symmetric product orbifold points are located in moduli space. 
The symmetric product orbifold features a higher spin symmetry \cite{Gaberdiel:2015wpo, Gaberdiel:2014cha, Eberhardt:2018plx}, whose emergence has so far not been completely elucidated from a string theory point of view \cite{Sundborg:2000wp} (see \cite{Ferreira:2017pgt, Gaberdiel:2018rqv, Giribet:2018ada} for recent observations concerning the emergence of this higher spin symmetry in pure NS-NS background for $k=1$). 
We have so far not observed any additional massless fields in the spectrum.

The dispersion relation obtained using integrability methods \cite{Hoare:2013lja}, in the decompactification limit, contains a term which is linear in the mode number and a squared sine term. 
In \cite{Hofman:2006xt} the comparison with the giant magnon solution suggested a transcendental analytic structure of the string spectrum. 
On the other hand, our conformal weights arise through diagonalisation, so the current algebra approach we have presented can only produce an algebraic structure for the spectrum. 
In particular, we cannot reproduce the giant magnon solution. 
We believe this is not a contradiction: the giant magnon solution is not physical, since it is not level-matched, and likewise we can so far only compute non level-matched conformal weights, as mentioned above. 
So there is no a priori reasons for the two formulas to agree. 
It would be very interesting to establish a connection between the two approaches.
%%%%%%%%%%%%%%%%%%%%%%%%%%%%%%
\section*{Acknowledgements}
%%%%%%%%%%%%%%%%%%%%%%%%%%%%%%
It is a pleasure to thank Andrea Dei, Matthias Gaberdiel, Shota Komatsu, Juan Maldacena, Alessandro Sfondrini, Edward Witten, and Ida Zadeh for useful conversations.
We would like to thank Matthias Gaberdiel for guidance during this work, for initial collaboration and for comments on the manuscript.
LE thanks the Institute for Advanced Study in Princeton and Stockholm University for hospitality while the bulk of this work was done. Both our research is supported by the NCCR SwissMAP, funded by the Swiss National Science Foundation.

\appendix
%%%%%%%%%%%%%%%%%%%%%%%%%%%%%%%%%%%%%%%%%%%%%%%%%%%%%%%%%%%%%%%%%%%%%%%
\section{Properties of Lie superalgebras} \label{app:properties}
%%%%%%%%%%%%%%%%%%%%%%%%%%%%%%%%%%%%%%%%%%%%%%%%%%%%%%%%%%%%%%%%%%%%%%%

%%%%%%%%%%%%%%%%%%%%%%%%%%%%%%%%%%%%%%%%%%%%%%%%%%%%%%%%%%%%%%%%%%%%%%%
\subsection{Root system and classification}
%%%%%%%%%%%%%%%%%%%%%%%%%%%%%%%%%%%%%%%%%%%%%%%%%%%%%%%%%%%%%%%%%%%%%%%
In this paper, basic classical Lie superalgebras with vanishing dual Coxeter number play an important r\^ole. 
These are completely classified by Ka\v c \cite{Kac:1977em} and are \cite{Quella:2013oda, Frappat:1987ix, Frappat:1996pb}:
\be 
\mathfrak{psl}(n|n)\, , \qquad \mathfrak{osp}(2n+2|2n)\qquad\text{and}\qquad \mathfrak{d}(2,1;\alpha)\, ,
\ee
where $n \ge 1$ is any integer and $\alpha>1$.\footnote{We have the isomorphism $\mathfrak{osp}(4|2)\cong \mathfrak{d}(2,1;\alpha=1)$ and $\mathfrak{d}(2,1;\alpha) \cong \mathfrak{d}(2,1;\alpha^{-1})$ for $\alpha \in \mathds{R}$. Since we want to choose a real form, we also restrict to $\alpha \in \mathds{R}$. We also have an isomorphism $\mathfrak{d}(2,1;\alpha\to \infty)\cong\mathfrak{psu}(1,1|2) \rtimes \mathfrak{su}(2)$, where $\mathfrak{su}(2)$ acts as an outer automorphism on $\mathfrak{psu}(1,1|2)$. This will discussed further in Appendix~\ref{subapp:d21alpha}.} 
These algebras are called basic because they possess an invariant bilinear form, which we need to construct the action, and hence also the Virasoro tensor.
For simple superalgebras a Cartan subalgebra $\mathfrak{h}$ can be chosen. 
For basic Lie superalgebras $\mathfrak{h}$ agrees with the Cartan subalgebra of the bosonic subalgebra, and is thus unique up to conjugation. 
Hence a root system can be defined.% of which we make some use in the main part of the work.

It is well-known (and explained in the Section~\ref{sec:current}) that the principal model with WZW-term on a supergroup is a CFT if the respective dual Coxeter number vanishes. The dual Coxeter number is half of the Casimir of the adjoint representation, so in particular the quadratic Casimir of the adjoint representation vanishes for the above Lie superalgebras.

%%%%%%%%%%%%%%%%%%%%%%%%%%%%%%%%%%%%%%%%%%%%%%%%%%%%%%%%%%%%%%%%%%%%%%%
\subsection{The (affine) Lie superalgebra $\mathfrak{psu}(1,1|2)$} \label{subapp:psu112}
%%%%%%%%%%%%%%%%%%%%%%%%%%%%%%%%%%%%%%%%%%%%%%%%%%%%%%%%%%%%%%%%%%%%%%%%

The algebra $\mathfrak{psu}(1,1|2)$ plays an important r\^ole when applying our formalism to string theory, so we recall here the relevant commutation relations. 
We display here the commutation relations for the affine algebra $\mathfrak{psu}(1,1|2)_k$.
The commutation relations for the global algebra follow by looking at the zero-modes only. 
We use a spinor notation for the algebra. 
In particular, the indices $\alpha,\beta,\gamma$ denote spinor indices and take values $\{\pm\}$. 
The bosonic subalgebra of $\mathfrak{psu}(1,1|2)_k$ consists of $\mathfrak{sl}(2,\mathds{R})_k \oplus \mathfrak{su}(2)_k\cong \mathfrak{su}(2)_{-k} \oplus \mathfrak{su}(2)_k$, whose modes we denote by $J^a_m$ and $K^a_m$, respectively. 
The fermionic generators are denoted $S^{\alpha\beta\gamma}_n$.
They satisfy the commutation relations \cite{Gotz:2006qp, Gaberdiel:2011vf}:\footnote{Note that these conventions differ from the ones usually employed in the RNS formalism for $\mathfrak{sl}(2,\mathds{R})$, see e.g.~\cite{Ferreira:2017pgt}.}
\begin{align}
\begin{aligned}
\, [J^3_m,J^3_n]&=-\tfrac{1}{2}km\delta_{m+n,0}\, , & [K^3_m,K^3_n]&=\tfrac{1}{2}km\delta_{m+n,0}\, , \\
[J^3_m,J^\pm_n]&=\pm J^\pm_{m+n}\, , &[K^3_m,K^\pm_n]&=\pm K^\pm_{m+n}\, , \\
[J^+_m,J^-_n]&=-km\delta_{m+n,0}+2J^3_{m+n}\,,\hspace{-.29cm} & [K^+_m,K^-_n]&=km\delta_{m+n,0}+2K^3_{m+n}\, , \\
[J^a_m,S^{\alpha\beta\gamma}_n]&=\tfrac{1}{2}\tensor{(\sigma^a)}{^\alpha_\mu} S^{\mu\beta\gamma}_{m+n}\, , & [K^a_m,S^{\alpha\beta\gamma}_n]&=\tfrac{1}{2}\tensor{(\sigma^a)}{^\beta_\nu} S^{\alpha\nu\gamma}_{m+n}\, , \\
 \{S^{\alpha\beta\gamma}_m,S^{\mu\nu\rho}_n\}&=km \epsilon^{\alpha\mu}\epsilon^{\beta\nu}\epsilon^{\gamma\rho}\delta_{m+n,0}-\epsilon^{\beta\nu}\epsilon^{\gamma\rho} \tensor{(\sigma_a)}{^{\alpha\mu}} J^a_{m+n}+\epsilon^{\alpha\mu}\epsilon^{\gamma\rho} \tensor{(\sigma_a)}{^{\beta\nu}} K^a_{m+n}\,.\hspace{-20cm}
\end{aligned}
\end{align}
Here $a \in \{\pm,3\}$ denote adjoint indices of $\mathfrak{su}(2)$ or $\mathfrak{sl}(2,\mathds{R})$. 
The two Cartan generators are chosen to be $J^3_0$ and $K^3_0$, and we denote their eigenvalues throughout the text as $j$ and $\ell$, respectively. 
Furthermore, there is a unique (up to rescaling) invariant form on $\mathfrak{psu}(1,1|2)$, which can be read off from the central terms:
\begin{align} 
\begin{aligned}
\kappa(J^3,J^3)&=-\tfrac{1}{2}\,, \ \kappa(J^\pm,J^\mp)=-1\,,\ \kappa(K^3,K^3)=\tfrac{1}{2}\,, \ \kappa(K^\pm,K^\mp)=1\,,\\
\kappa(S^{\alpha\beta\gamma},S^{\mu\nu\rho})&=\epsilon^{\alpha\mu}\epsilon^{\beta\nu}\epsilon^{\gamma\rho} \, ,\ \kappa(S^{\alpha\beta\gamma},J^a)=0\,,\ \kappa(S^{\alpha\beta\gamma},K^a)=0\,.
\end{aligned}
\end{align}

The representations we will consider for string theory applications are lowest weight for the $\mathfrak{sl}(2,\mathds{R})$-oscillators, and half-infinite.\footnote{Also so-called spectrally flowed representations occur. For non-vanishing R-R-flux, they cannot be described on the level of the algebra $\mathfrak{psu}(1,1|2)_k$ alone.}  For $\mathfrak{su}(2)$, they are finite dimensional. Hence they are characterized by
\begin{align}
\begin{aligned}
J^3_0 |j,\ell \rangle&=j |j,\ell \rangle\ ,& K^3_0 |j,\ell \rangle &= \ell |j,\ell \rangle\, , \\
J^-_0 |j,\ell\rangle&=0\ , & K^+_0 |j,\ell \rangle&=0\, , \\
J^a_m |j,\ell \rangle&=0\ , \quad m>0\ , & K^a_m |j,\ell \rangle &=0 \ , \quad m>0\, .
\end{aligned} \label{eq:psu112 highest weight condition}
\end{align}
Requiring that the zero-mode representation has no negative-norm states imposes $\ell \in \tfrac{1}{2}\mathds{Z}_{\ge 0}$, $j$ is continuous.
The Casimir of such a representation reads
\be 
\mathcal{C}(j,\ell)=-2j(j-1)+2\ell(\ell+1)\, . \label{eq:psu112 casimir} 
\ee
A representation $\vert j,\ell\rangle$ is atypical if the BPS bound $j\ge \ell+1$ is saturated, and it is otherwise typical.
A typical representation $\vert j,\ell \rangle$ consists of the following 16 $\mathfrak{sl}(2,\mathds{R})\oplus \mathfrak{su}(2)$-multiplets:
\be 
4(j,\ell)\,,\ (j\pm 1,\ell)\,,\ (j,\ell \pm 1)\,,\ 2(j \pm\tfrac{1}{2},\ell\pm \tfrac{1}{2})\, . \label{eq:psu112 multiplet}
\ee
%%%%%%%%%%%%%%%%%%%%%%%%%%%%%%%%%%%%%%%%%%%%%%%%%%%%%%%%%%%%%%%%%%%%%%%
\subsection{The (affine) Lie superalgebra $\mathfrak{d}(2,1;\alpha)$} \label{subapp:d21alpha}
%%%%%%%%%%%%%%%%%%%%%%%%%%%%%%%%%%%%%%%%%%%%%%%%%%%%%%%%%%%%%%%%%%%%%%%% 
The (affine) Lie superalgebra $\mathfrak{d}(2,1;\alpha)$ is used to describe string theory on the background ${\rm AdS}_3 \times {\rm S}^3 \times{\rm S}^3 \times {\rm S}^1$, so we review it here and fix our conventions. The non-vanishing commutation relations for the affine algebra take the form
\begin{align}
\begin{aligned}
\, [J^3_m,J^3_n]&=-\tfrac{1}{2}km\delta_{m+n,0}\, , & [K^{(\pm)3}_m,K^{(\pm)3}_n]&=\tfrac{1}{2}k^\pm m\delta_{m+n,0}\, , \\
[J^3_m,J^\pm_n]&=\pm J^\pm_{m+n}\, , &[K^{(\pm)3}_m,K^{(\pm)\pm}_n]&=\pm K^{(\pm)\pm}_{m+n}\, , \\
[J^+_m,J^-_n]&=-km\delta_{m+n,0}+2J^3_{m+n}\,, & [K^{(\pm)+}_m,K^{(\pm)-}_n]&=k^\pm m\delta_{m+n,0}+2K^{(\pm)3}_{m+n}\, , \hspace{-3cm}\\
[J^a_m,S^{\alpha\beta\gamma}_n]&=\tfrac{1}{2}\tensor{(\sigma^a)}{^\alpha_\mu} S^{\mu\beta\gamma}_{m+n}\, , & [K^{(+)a}_m,S^{\alpha\beta\gamma}_n]&=\tfrac{1}{2}\tensor{(\sigma^a)}{^\beta_\nu} S^{\alpha\nu\gamma}_{m+n}\, , \\
[K^{(-)a}_m,S^{\alpha\beta\gamma}_n]&=\tfrac{1}{2}\tensor{(\sigma^a)}{^\gamma_\rho} S^{\alpha\beta\rho}_{m+n}\,\\
 \{S^{\alpha\beta\gamma}_m,S^{\mu\nu\rho}_n\}&=km \epsilon^{\alpha\mu}\epsilon^{\beta\nu}\epsilon^{\gamma\rho}\delta_{m+n,0}-\epsilon^{\beta\nu}\epsilon^{\gamma\rho} \tensor{(\sigma_a)}{^{\alpha\mu}} J^a_{m+n}+\gamma\epsilon^{\alpha\mu}\epsilon^{\gamma\rho} \tensor{(\sigma_a)}{^{\beta\nu}} K^{(+)a}_{m+n}\hspace{-20cm}\\
 &\qquad+(1-\gamma)\epsilon^{\alpha\mu}\epsilon^{\beta\nu} \tensor{(\sigma_a)}{^{\gamma\rho}} K^{(-)a}_{m+n}\,.\hspace{-20cm}
\end{aligned}
\end{align}
Again, $\alpha,\beta,\dots$ are spinor indices and take values in $\{\pm\}$. As before, $a$ is an adjoint index and takes values in $\{\pm,3\}$. It is raised an lowered by the standard $\mathfrak{su}(2)$-invariant form. Finally, $\gamma$, $k^+$ and $k^-$ are related to $\alpha$ and $k$ by
\be 
\gamma=\frac{\alpha}{1+\alpha}\, , \qquad k^+=\frac{(\alpha+1)k}{\alpha} \, , \qquad k^-=(\alpha+1)k\, .
\ee
We note that $k^+,k^- \in \mathds{Z}_{\ge 0}$, so the affine algebra imposes furthermore $\alpha \in \mathds{Q}_{\ge 0} \cup \{\infty\}$.

In the limit $\alpha \to \infty$, $\gamma=1$ and the modes $K^{(-)a}_m$ decouple from the rest of the algebra. After decoupling, the algebra becomes again $\mathfrak{psu}(1,1|2)_k$.\footnote{Physically, this corresponds to the fact that when decompactifying one of the three-spheres, the geometry of the background becomes $\mathrm{AdS}_3 \times \mathrm{S}^3 \times \mathds{R}^3 \times \mathrm{S}^1$, which is locally isometric to $\mathrm{AdS}_3 \times \mathrm{S}^3 \times\mathbb{T}^4$ and hence has $\mathfrak{psu}(1,1|2)$ as a symmetry algebra.}
There is a unique (up to rescaling) invariant form, which can be read off from the central terms in the commutation relations.

We will consider representations which are half-infinite in the $\mathfrak{sl}(2,\mathds{R})$, and finite-dimensional in the two $\mathfrak{su}(2)$'s. Hence they are again characterized by \eqref{eq:psu112 highest weight condition}, where the conditions on the $K$-modes apply to both $K^{(+)a}_m$ and $K^{(-)a}_m$. A representation is consequently parametrized by the three spins $|j,\ell^+,\ell^-\rangle$. 
A representation $\vert j,\ell^+,\ell^-\rangle$ is atypical if the BPS bound
\be\label{eq: d21alpha BPS}
j\ge \gamma\ell^+ + (1-\gamma)\ell^- \, ,
\ee
is saturated, and it is otherwise typical.
A typical multiplet consists of the following 16 $\mathfrak{sl}(2,\mathds{R}) \oplus \mathfrak{su}(2) \oplus \mathfrak{su}(2)$-representations:
\begin{multline}
2(j,\ell^+,\ell^-)\,,\ (j\pm 1,\ell^+,\ell^-)\,,\ (j,\ell^+\pm 1,\ell^-)\,,\\ (j,\ell^+,\ell^-\pm 1)\,, \ (j \pm \tfrac{1}{2},\ell^+\pm \tfrac{1}{2},\ell^-\pm \tfrac{1}{2})\, .
\end{multline}
Its quadratic Casimir reads
\be 
{\cal C}(j,\ell^+,\ell^-)=-2j(j-1)+2\gamma \ell^+(\ell^++1)+2(1-\gamma) \ell^-(\ell^-+1)\, .
\ee
In the main text, we shall find it useful to parametrize $\gamma$ by the angle $\varphi$ through $\gamma=\cos^2\varphi$.

\bibliographystyle{JHEP}
\bibliography{BMNbib}

\end{document}